\magnification=1200
\hsize=6.5 truein
\vsize=9.0 truein
\font\tencaps   = cmcsc10                
\font\eightit   = cmti8                  
\def\singlespace{\baselineskip 12pt \lineskip 1pt \parskip 2pt plus 1 pt}

\def\today{\number\day\enspace
     \ifcase\month\or January\or Febuary\or March\or April\or May\or
     June\or July\or August\or September\or October\or
     November\or December\fi \enspace\number\year}
\def\clock{\count0=\time \divide\count0 by 60
    \count1=\count0 \multiply\count1 by -60 \advance\count1 by \time
    \number\count0:\ifnum\count1<10{0\number\count1}\else\number\cofi}
\footline={\hss -- \folio\ -- \hss}

\def\kms{km s$^{-1}$}
\def\deg{\ifmmode^\circ\;\else$^\circ\;$\fi}
\def\solar{\ifmmode_{\mathord\odot}\;\else$_{\mathord\odot}\;$\fi}
\def\amin{\ifmmode^\prime\kern- .38em\;\else$^\prime\kern- .38em\;$\fi}
\def\asec{\ifmmode^{\prime\prime}\kern- .38em\;\else$^{\prime\prime}
    \kern- .38em\;$\fi}
\def\s.{\kern+ .1em\lower 0.0ex\hbox{$\buildrel ^{\prime\prime} \over 
    {\rm .} \kern- .08em$}} 
\def\m.{\kern+ .1em\lower 0.0ex\hbox{$\buildrel ^{\prime} \over 
    {\rm .} \kern- .08em$}} 
\def\grad{\kern+ .1em\lower 0.0ex\hbox{$\buildrel {\rightharpoonup} \over 
    {\nabla} \kern- .08em$}} 
\def\div{\kern+ .1em\lower 0.0ex\hbox{$\buildrel {\rightharpoonup} \over 
    {\nabla} \kern- .08em$}\cdot} 
\def\ltsim{\lower 0.5ex\hbox{$\; \buildrel < \over \sim \;$}} 
\def\gtsim{\lower 0.5ex\hbox{$\; \buildrel > \over \sim \;$}} 
\def\x{\enspace}

\def\et{{\it et~al.~}}

\def\ts{\thinspace}  
\def\jref#1 #2 #3 #4 {{\par\noindent \hangindent=3em \hangafter=1 
      \advance \rightskip by 5em #1, {\it#2}, {\bf#3}, #4\par}}
\def\ref#1{{\par\noindent \hangindent=3em \hangafter=1 
      \advance \rightskip by 5em #1\par}}

\newcount\eqnum
\def\nexteq{\global\advance\eqnum by1 \eqno(\number\eqnum)}
\def\lasteq#1{\if)#1[\number\eqnum]\else(\number\eqnum)\fi#1}
\def\preveq#1#2{{\advance\eqnum by-#1
    \if)#2[\number\eqnum]\else(\number\eqnum)\fi}#2}

\def\tableheight{\vrule width 0pt height 8.5pt depth 3.5pt}
{\catcode`|=\active \catcode`&=\active 
    \gdef\tabledelim{\catcode`|=\active \let|=\vbar
                     \catcode`&=\active \let&=\nobar} }
\def\table{\begingroup
    \def\twidth{\hsize}
    \def\tablewidth##1{\def\twidth{##1}}
    \def\defaultheight{\vrule width 0pt height 8.5pt depth 3.5pt}
    \def\heightdepth##1{\dimen0=##1
        \ifdim\dimen0>5pt 
            \divide\dimen0 by 2 \advance\dimen0 by 2.5pt
            \dimen1=\dimen0 \advance\dimen1 by -5pt
            \vrule width 0pt height \the\dimen0  depth \the\dimen1
        \else  \divide\dimen0 by 2
            \vrule width 0pt height \the\dimen0  depth \the\dimen0 \fi}
    \def\spacing##1{\def\defaultheight{\heightdepth{##1}}}
    \def\nextheight##1{\noalign{\gdef\tableheight{\heightdepth{##1}}}}
    \def\end{\cr\noalign{\gdef\tableheight{\defaultheight}}}
    \def\zerowidth##1{\omit\hidewidth ##1 \hidewidth}    
    \def\hline{\noalign{\hrule}}
    \def\dline{\noalign{\hrule \vskip 2pt \hrule}}
    \def\skip##1{\noalign{\vskip##1}}
    \def\bskip##1{\noalign{\hbox to \twidth{\vrule height##1 depth 0pt \hfil
        \vrule height##1 depth 0pt}}}
    \def\header##1{\noalign{\hbox to \twidth{\hfil ##1 \unskip\hfil}}}
    \def\bheader##1{\noalign{\hbox to \twidth{\vrule\hfil ##1 
        \unskip\hfil\vrule}}}
    \def\spanloop{\span\omit \advance\mscount by -1}
    \def\extend##1##2{\omit
        \mscount=##1 \multiply\mscount by 2 \advance\mscount by -1
        \loop\ifnum\mscount>1 \spanloop\repeat \ \hfil ##2 \unskip\hfil}
    \def\vbar{&\vrule&}
    \def\nobar{&&}
    \def\hdash##1{ \noalign{ \relax \gdef\tableheight{\heightdepth{0pt}}
        \toks0={} \count0=1 \count1=0 \putout##1\end 
        \toks0=\expandafter{\the\toks0 &\end} \xdef\piggy{\the\toks0} }
        \piggy}
    \let\e=\expandafter
    \def\putspace{\ifnum\count0>1 \advance\count0 by -1
        \toks0=\e\e\e{\the\e\toks0\e&\e\multispan\e{\the\count0}\hfill} 
        \fi \count0=0 }
    \def\putrule{\ifnum\count1>0 \advance\count1 by 1
        \toks0=\e\e\e{\the\e\toks0\e&\e\multispan\e{\the\count1}\leaders\hrule\hfill}
        \fi \count1=0 }
    \def\putout##1{\ifx##1\end \putspace \putrule \let\next=\relax 
        \else \let\next=\putout
            \ifx##1- \advance\count1 by 2 \putspace
            \else    \advance\count0 by 2 \putrule \fi \fi \next}   }
\def\tablespec#1{
    \def\vdimens{\noexpand\tableheight}
    \def\tabby{\tabskip=0pt plus100pt minus100pt}
    \def\r{&################\tabby&\hfil################\unskip}
    \def\c{&################\tabby&\hfil################\unskip\hfil}
    \def\l{&################\tabby&################\unskip\hfil}
    \edef\templ{\noexpand\vdimens ########\unskip  #1 
         \unskip&########\tabskip=0pt&########\cr}
    \tabledelim
    \edef\body##1{ \vbox{
        \tabskip=0pt \offinterlineskip
        \halign to \twidth {\templ ##1}}} }
\rm
\singlespace
\raggedbottom

\centerline{\bf THE CENTERS OF EARLY-TYPE GALAXIES WITH {\it HST}. \x IV.}
\centerline{{\bf CENTRAL PARAMETER RELATIONS}\footnote{$^1$}{Based on 
observations with the NASA/ESA {\it Hubble Space Telescope},
obtained at the Space Telescope Science Institute, which is operated by AURA,
Inc., under NASA contract NAS 5-26555.}  }

\bigskip
\centerline{\tencaps S.M. Faber}
\centerline{UCO/Lick Observatory, Board of Studies in 
Astronomy and Astrophysics,}
\centerline{ University of California, Santa Cruz, CA 95064}
\centerline{Electronic mail: faber@ucolick.org}

\medskip
\centerline{\tencaps Scott Tremaine}
\centerline{CIAR Cosmology and Gravity Program, Canadian Institute for Theoretical
Astrophysics} 
\centerline{University of Toronto, 60 St. George St., Toronto M5S 3H8, Canada}
\centerline{Electronic mail: tremaine@cita.utoronto.ca}

\medskip
\centerline{\tencaps Edward A. Ajhar}
\centerline{Kitt Peak National Observatory, National Optical 
Astronomy Observatories\footnote{$^2$}{Operated by AURA under 
cooperative agreement with the U. S. National Science Foundation.},}
\centerline{P.O. Box 26732, Tucson, AZ 85726}
\centerline{Electronic mail: ajhar@noao.edu}

\medskip
\centerline{{\tencaps Yong-Ik Byun}\footnote{$^3$}{Present address: 
Institute of
Astronomy, National Central University, Chung-Li, Taiwan 32054, R. O. C.}   }
\centerline{Institute for Astronomy, University of Hawaii, 
2680 Woodlawn Dr., Honolulu, HI 96822}
\centerline{Electronic mail: byun@haneul.phy.ncu.edu.tw}

\medskip
\centerline{\tencaps Alan Dressler}
\centerline{The Observatories of the Carnegie Institution, 
813 Santa Barbara St., Pasadena, CA 91101}
\centerline{Electronic mail: dressler@ociw.edu}

\medskip
\centerline{{\tencaps Karl Gebhardt}\footnote{$^4$}{Present address: 
UCO/Lick Observatory, University of California, Santa Cruz, CA 95064.}}
\centerline{Department of Astronomy, University of Michigan, 
Ann Arbor, MI 48109}
\centerline{Electronic mail: gebhardt@astro.lsa.umich.edu}

\medskip
\centerline{{\tencaps Carl Grillmair}\footnote{$^5$}{Present address: Jet
Propulsion Laboratory, Mail Stop 183-900, 4800 Oak Grove Drive, Pasadena, 
CA 91109.}}
\centerline{UCO/Lick Observatory, Board of Studies in 
Astronomy and Astrophysics,}
\centerline{ University of California, Santa Cruz, CA 95064}
\centerline{Electronic mail: carl@grandpa.jpl.nasa.gov}

\vfill\eject
\medskip
\centerline{\tencaps John Kormendy}
\centerline{Institute for Astronomy, 
University of Hawaii, 2680 Woodlawn Dr., Honolulu, HI 96822}
\centerline{Electronic mail: kormendy@oort.ifa.hawaii.edu}

\medskip
\centerline{\tencaps Tod R. Lauer}
\centerline{Kitt Peak National Observatory, National Optical 
Astronomy Observatories$^2$,}
\centerline{P.O. Box 26732, Tucson, AZ 85726}
\centerline{Electronic mail: lauer@noao.edu}

\medskip
\centerline{\tencaps Douglas Richstone}
\centerline{Department of Astronomy, University of Michigan, 
Ann Arbor, MI 48109}
\centerline{Electronic mail: dor@astro.lsa.umich.edu}

\bigskip\bigskip\bigskip\bigskip
\centerline{\it Received
\_ \hskip -0.1 truein \_ \hskip -0.1 truein \_ \hskip -0.1 truein \_ \hskip -0.1 truein \_ \hskip -0.1 truein 
\_ \hskip -0.1 truein \_ \hskip -0.1 truein \_ \hskip -0.1 truein \_ \hskip -0.1 truein \_ \hskip -0.1 truein 
\_ \hskip -0.1 truein \_ \hskip -0.1 truein \_ \hskip -0.1 truein \_ \hskip -0.1 truein \_ \hskip -0.1 truein 
\_ \hskip -0.1 truein \_ \hskip -0.1 truein \_ \hskip -0.1 truein \_ \hskip -0.1 truein \_ \hskip -0.1 truein 
\_ \hskip -0.1 truein \_ \hskip -0.1 truein \_ \hskip -0.1 truein \_ \hskip -0.1 truein \_ \hskip -0.1 truein 
\_ \hskip -0.1 truein \_ \hskip -0.1 truein \_ \hskip -0.1 truein \_ \hskip -0.1 truein \_ \hskip -0.1 truein 
\_ \hskip -0.1 truein \_ \hskip -0.1 truein \_ \hskip -0.1 truein \_ \hskip -0.1 truein \_ \hskip -0.1 truein 
\_ \hskip -0.1 truein \_ \hskip -0.1 truein \_ \hskip -0.1 truein \_ \hskip -0.1 truein \_ \hskip -0.1 truein 
\x ; 
Revised
\_ \hskip -0.1 truein \_ \hskip -0.1 truein \_ \hskip -0.1 truein \_ \hskip -0.1 truein \_ \hskip -0.1 truein 
\_ \hskip -0.1 truein \_ \hskip -0.1 truein \_ \hskip -0.1 truein \_ \hskip -0.1 truein \_ \hskip -0.1 truein 
\_ \hskip -0.1 truein \_ \hskip -0.1 truein \_ \hskip -0.1 truein \_ \hskip -0.1 truein \_ \hskip -0.1 truein 
\_ \hskip -0.1 truein \_ \hskip -0.1 truein \_ \hskip -0.1 truein \_ \hskip -0.1 truein \_ \hskip -0.1 truein 
\_ \hskip -0.1 truein \_ \hskip -0.1 truein \_ \hskip -0.1 truein \_ \hskip -0.1 truein \_ \hskip -0.1 truein 
\_ \hskip -0.1 truein \_ \hskip -0.1 truein \_ \hskip -0.1 truein \_ \hskip -0.1 truein \_ \hskip -0.1 truein 
\_ \hskip -0.1 truein \_ \hskip -0.1 truein \_ \hskip -0.1 truein \_ \hskip -0.1 truein \_ \hskip -0.1 truein 
\_ \hskip -0.1 truein \_ \hskip -0.1 truein \_ \hskip -0.1 truein \_ \hskip -0.1 truein \_ \hskip -0.1 truein}

\vfill\eject

\centerline{ABSTRACT}

\bigskip
We analyze {\it Hubble Space Telescope} surface-brightness profiles of 61
elliptical galaxies and spiral bulges (hereafter ``hot" galaxies).  The
profiles are parameterized by break radius $r_b$ and break surface brightness
$I_b$.  These are combined with central velocity dispersions, total
luminosities, rotation velocities, and isophote shapes to explore correlations
among central and global properties.  Luminous hot galaxies ($M_V < -22$)
have {\it cuspy cores} with steep outer power-law profiles that
break at $r \approx r_b$ to shallow inner profiles $I \propto r^{-\gamma}$
with $ \gamma \le 0.3$.
Break radii and core luminosities for these objects are approximately
proportional to effective radii and total
luminosities.  Scaling relations are presented for
several core parameters as a function of total luminosity.
Cores follow a fundamental plane that parallels the global fundamental
plane for hot galaxies but is 30\% thicker.  Some of this extra thickness may
be due to the effect of massive black holes (BHs) on central velocity
dispersions.  Faint hot galaxies ($M_V > -20.5$) show steep, largely
featureless {\it power-law} profiles that lack cores.  Measured values of
$r_b$ and $I_b$ for these galaxies are limits only.  
At a limiting radius of 10 pc, the centers of power-law
galaxies are up to 1000 times denser in mass and luminosity than the
cores of large galaxies.  At intermediate
magnitudes ($-22 < M_V < -20.5$), core and power-law galaxies coexist, and
there is a range in $r_b$ at a given luminosity of at least two orders of
magnitude.  Here, central properties correlate strongly with global rotation and
shape: core galaxies tend to be boxy and slowly rotating, whereas power-law
galaxies tend to be disky and rapidly rotating.  
A search for inner disks was conducted to test a
claim in the literature, based on a smaller sample, that power laws originate
from edge-on stellar disks.  We find only limited evidence for such disks and
believe that the difference between core and power-law profiles 
reflects a
real difference in the spatial distribution of the luminous {\it spheroidal}
component of the galaxy. The dense power-law centers of disky, rotating
galaxies are consistent with their formation in gas-rich mergers. The
parallel proposition, that cores are the by-products of gas-free stellar mergers, is
less compelling for at least two reasons: (1) dissipationless hierarchical
clustering does not appear to produce core profiles like those seen; (2) core
galaxies accrete small, dense, gas-free galaxies at a rate sufficient to fill in
their low-density cores if the satellites survived and sank to the center (whether
the satellites survive is still an open question). An alternative
model for core formation involves 
the orbital decay of massive black holes (BHs) that are 
accreted in mergers: the decaying BHs may heat and
eject stars from the center, eroding a power law if any exists and
scouring out a core. An average BH mass per spheroid of 0.002 times the
stellar mass yields cores in fair agreement with 
observed cores and is consistent with the energetics of AGNs and
the kinematic detection of BHs in nearby galaxies. An unresolved issue is why
power-law galaxies also do not have cores if this process operates in all hot
galaxies. 

\vfill\eject

\centerline{1. INTRODUCTION}
\bigskip

The {\it Hubble Space Telescope} (HST) allows us to study the centers of
nearby galaxies with a resolution of a few parsecs.  The centers of galaxies
are interesting for several reasons: (1) some galaxy centers harbor AGNs and
QSOs; (2) many or most galaxy centers may contain
massive black holes that are the remnants of dead QSOs; (3) dynamical
processes such as relaxation are more rapid near galaxy centers than elsewhere
in the galaxy; thus interesting dynamical phenomena are likely to occur first
near the center; (4) galaxy centers are to galactic astronomy as middens are
to archaeologists: centers are the bottoms of potential
wells and debris such as gas and dense stellar systems settle there,
providing a record of the past history of the galaxy.

The systematic properties of the centers of ellipticals and spiral bulges
(hereafter ``hot galaxies") were described by Lauer (1983, 1985a) and Kormendy
(1982a, 1984, 1985, 1987a,b).  They detected inner regions in many galaxies
where the slope of the surface-brightness profile flattens out, which they
termed {\it cores}.  They measured the size and surface brightness of these
cores and demonstrated {\it central parameter relations} that linked core
properties with one another and with global properties such as luminosity and
effective radius. In particular, cores in brighter galaxies
were larger and of lower density.  The most recent version of the central
parameter relations using ground-based data was presented by Kormendy and
McClure (1993). A major goal of this paper is to revisit the central parameter
relations using new {\it HST} data on 61 galaxies.  We shall show that {\it
HST} broadly supports the ground-based scaling relations but elaborates upon
them in important ways.

Historically, the existence of cores in hot galaxies has been accepted as
``normal'' --- probably because familiar dynamical models for galaxies
such as the isothermal sphere and King models possess cores.  In the absence
of a central compact mass, it is plausible that all physical variables should
vary smoothly near the origin and hence be expandable in a Taylor series with
only even powers of $r$.  In particular, the surface brightness may be written
$$ I(r) = I_0 + I_1r^2 + \hbox{O}(r^4), \eqno{(1)} $$
where $r$ is projected radius. Using the conventional definition of core
radius, a galaxy satisfying Eq. (1) would exhibit a core of radius $r_c$
such that $I(r_c) = {1 \over 2}I(0)$. Tremaine (1997) suggests the term
``analytic core'' for systems with cores in which all physical variables vary
smoothly. The cores of King models and the isothermal sphere are thus
analytic, while the $R^{1/4}$ law is not.

{\it HST} observations show that real cores are not analytic.  In analytic
cores, the surface brightness flattens at small radii as $d\log I/d\log r
\propto r^2$ --- note that this is stronger than the usual condition for a flat
profile, $d\log I/d\log r \to 0$ --- whereas real cores show shallow power-law
cusps into the resolution limit (Crane \et 1993; 
Kormendy \et 1994; Jaffe \et 1994; Lauer \et 1995, hereafter
Paper I; Kormendy \et 1996a).  Fits in Byun \et (1996, Paper II) yield projected slopes
$\gamma \equiv -d\log I/d\log r$ in the range 0.05--0.3 for surface
brightness, while non-parametric inversions for space density 
show even steeper slopes, from 0.2 to 1.1 (Gebhardt
\et 1996, Paper III; and Kormendy \et 1996a).  Thus, real cores
have divergent rather than constant densities as $r\to0$.

So far, cores have been found only in luminous ellipticals.  The division
between core and non-core galaxies is fairly sharp. Surface-brightness
profiles either flatten out to form cores or continue to rise steeply into the
resolution limit --- few galaxies are in between (Kormendy \et 1994; Jaffe \et
1994; Paper I; Kormendy \et 1996a).  Statistical analysis of 
non-parametrically derived space density profiles 
indicates the existence of two groups (core and non-core) at the 90\%
confidence level (Paper III).  

Paper I introduced the term {\it power laws} to describe
the steeply rising, featureless profiles that lack
cores.\footnote{$^5$} {Cores and power laws were also identified 
by Jaffe \et (1994), who called them Type I and Type II.}  
It is possible that the power-law category as we have drawn it may be
oversimplified:  At present the category contains a number of
low-luminosity galaxies whose upper limits on core size are 
larger than those predicted by extrapolation of
the core-luminosity relationship defined by brighter galaxies.
In other words, some of the low-luminosity power-law galaxies may
really be part of a core sequence extending to lower luminosity.
Recent WFPC2 images in fact show tiny cores in a few
power-law galaxies (Lauer \et 1997).  Nevertheless, the upper limits on core size for
{\it brighter} power-law galaxies are already well below the
core sequence for galaxies of similar luminosity, and thus clearly
differentiate them.   Future  results may  
compel some revision of the power-law category, but the present simple
core/power-law division is a useful working hypothesis.

Lauer (1985a)
emphasized that the {\it central} properties of hot galaxies do not correlate
perfectly with {\it total} luminosity and sought an explanation in terms of a
second parameter.  The present data suggest that this second parameter is
related to global rotation and isophote shape.  So far, cores have been found
only in luminous, slowly rotating ellipticals with boxy
isophotes\footnote{$^6$} {Isophote shape in elliptical galaxies is explained
and defined by Bender \& M\"ollenhoff (1987).  A recent discussion is given by
Kormendy \& Bender (1996).}, while power laws are found in faint, rapidly
rotating galaxies with disky isophotes. A link between central profile type
and global shape/rotation was suggested by Nieto
\et (1991a) based on ground-based images, and further evidence was presented
by Jaffe \et (1994) and Ferrarese \et (1994) based on {\it HST} images of 14
Virgo galaxies.  The present database is considerably 
larger and permits a critical 
examination of this link and its relation to hot galaxy formation.  Our point
of view differs importantly from that of Jaffe {\it et al.}, who ascribe
many of the differences between the two profile types to inclination effects
connected with a small inner disk seen either face-on or edge-on.  In
contrast, we --- like Nieto {\it et al.} --- believe that the spheroidal light
distributions are intrinsically different in the two types
and would look the same from any viewing angle.  These differences in viewpoint are
discussed in Section 5 and Appendix A.

The results we have described raise several theoretical issues: why are there
two types of profile and how did each type form?  Why do the two types
have different global rotation and shape?  Why are cores
non-analytic?  And what do central profiles tell us about hot galaxy
formation and evolution?

The second, more speculative, part of this paper addresses these
issues.
We suggest in Section 7 that the
power-law profiles of disky galaxies indicate dissipation and are therefore
consistent with formation in {\it gas-rich} mergers.  
The parallel suggestion ---
that the cores of boxy galaxies are the by-products of {\it purely stellar, 
gas-poor} mergers --- is more problematic.
For example, luminous core galaxies are
expected to accrete small dense
satellites.  The rate of such
accretions appears sufficient to gradually 
fill in all low-density cores if such 
satellites survived and sank to the center.  
An unresolved issue is whether the satellites do survive, and 
thus whether some other process is needed to 
defend low-density cores against in-fill.

Even if the data do not firmly require such a mechanism,
there is strong and growing evidence for a widespread population of massive
central black holes (BHs) in hot galaxies (Kormendy \& Richstone 1995). The
presence of these objects must be taken into account in standard 
merger-based models for
forming hot galaxies (Section 8).  The BHs associated with the merging
galaxies form binaries whose orbits then decay. The orbital decay heats the
surrounding stars, erodes a power law if one exists, and scours out a
core.  Accreted satellites will also tend to be ripped apart, thus preventing
core in-fill.
BHs with plausible masses (as estimated in Appendix B) seem able to produce cores
of roughly the right size and scaling versus galaxy
luminosity.  In this way, the presence of central BHs might
``rescue" the dissipationless, gas-poor model for cores and
boxy galaxies.  
However, models of core formation based purely on massive BHs leave
several questions open, notably how power-law profiles escape similar
disruption.

Whether or not these speculations about galaxy formation are correct, the
updated relations between central and global galaxy parameters that are
presented in this paper appear to provide important
new constraints on hot galaxy formation.

\bigskip\bigskip
\centerline{2. CENTRAL PROFILE TYPES}
\bigskip

Major collections of {\it HST} central profiles include Crane \et (1993),
Jaffe \et (1994), Forbes {\it et al.} (1995), and Paper I.  An assortment of
representative surface-bright\-ness profiles of 55 ellipticals and spiral
bulges is given in Fig. 1.  The following summary is based on the data
and discussion in Paper I.

We distinguish two types of hot galaxy:

\item{(1)} {\it Core galaxies} have ``broken" power-law surface-brightness
profiles that change slope significantly at a ``break radius" $r_b$.  To
identify a galaxy as having a core, we require that the absolute value of the
inner logarithmic slope, $\gamma \equiv -d\log I/d\log r$, be shallower than
0.3.  Nearly all core galaxies appear to have $\gamma > 0$, which indicates a
cusp in the central surface brightness and an even stronger cusp in the
luminosity density. Paper III concluded that, even with errors taken into account,
only 2 out of 15 known 
core galaxies could admit an analytic core ($\gamma = 0$).
Core galaxies as a class are luminous objects with $M_V \ltsim -20.5$ ($H_0=80$
km s$^{-1}$ Mpc$^{-1}$).  They range from brightest cluster galaxies down to
the intermediate-mass field elliptical NGC~3379.

\item{(2)} {\it Power-law galaxies} show fairly steep surface-brightness 
profiles with no significant break within 10\asec\ (at Virgo).  Their average
surface-brightness slope is $\gamma \simeq 0.8\pm0.3$ at the smallest
resolvable radius.  Power-law galaxies are generally fainter than core
galaxies ($M_V > -22$), but their luminosity densities at 10 pc are
10--1000 times higher than those of cores
(Paper I). Profile shapes within 0\s.1 are generally
not known, though recent WFPC2 images suggest small cores inside some power
laws.  Power-law galaxies include M~32 (NGC~221), small Virgo ellipticals, and
bulges of disk galaxies.

Both profile types are well fit by the following equation 
(the ``Nuker" law, Papers I and II):
$$ I(r) = I_b~2^{(\beta - \gamma)/\alpha}~
   {             \biggl({r_b \over r}\biggr)^{\gamma}  }
   {  \biggl[1 + \biggl({r \over r_b}\biggr)^{\alpha}\ts\biggr]^{(\gamma -
                                           \beta)/\alpha}  }.  \eqno{(2)}
$$
The asymptotic logarithmic slope inside $r_b$ is $-\gamma$, the asymptotic
outer slope is $-\beta$, and the parameter $\alpha$ parameterizes the sharpness of
the break.  The break radius $r_b$ is the point of maximum curvature in
log-log coordinates.  ``Break surface brightness," $I_b$, is the surface
brightness at $r_b$.  Equation (2) is intended to fit only over radii
accessible to the HST Planetary Camera, i.e., $<$10\asec.  For typical fitted
values of $\beta$, there must be a further turndown in the profile at larger
radii for the total luminosity to be finite.

{\it Nuclei} are identified when excess light above the prediction of Eq.  (2)
is visible within the inner few tenths of an arcsec.  Nuclei with varying
degrees of prominence are illustrated in Paper I (Fig. 14).  Objects with
prominent nuclei are always systems of low luminosity and are probably
nucleated dSph or dE galaxies.  Nuclei are assumed to be star clusters (or
possibly unresolved tiny stellar disks), but direct spectral confirmation is often
lacking.  A stellar nucleus in NGC~3115 has been resolved in recent WFPC2
images (Kormendy \et 1996b).  Non-thermal central point sources exist in four
galaxies in our sample: M~87 (NGC~4486), NGC~6166, Abell 2052 (Paper I) and
NGC~4594 (Kormendy
\et 1996c).  We call these {\it AGNs} to distinguish them from nuclei.  So
far, no nuclei (as opposed to AGNs) have been found within cores (Kormendy \&
Djorgovski 1989; Paper I).

Resolution plays an important role in classifying profiles and estimating
central properties.  This is illustrated in Fig. 2, which shows M~31 (NGC~224)
and M~32 (NGC~221) as seen at their actual distances and as they would be seen
24 times further away just beyond Virgo (for future reference, we call these
artificially positioned galaxies M~31-in-Virgo and M~32-in-Virgo). Up close,
M~31 shows a two-component profile that is clearly divided into a bulge and a
nucleus, the latter showing a small core.  The entire profile shows too much
substructure to fit comfortably into either the core or power-law category.
In contrast, M~31-in-Virgo shows only a trace of a nucleus, and its profile
and degree of nucleation are similar to those of several other galaxies that
we have classed as power laws (see Fig.  14 in Paper I for a collection of
power laws with varying degrees of nucleation).  M~31 implies that many
power-law galaxies, particularly those with hints of nuclei, may contain
significant substructure, including nuclei and tiny cores.

M~32 is similarly ambiguous. Seen up close, M~32's profile in Fig. 2 breaks
from a power law near 0\s.5, curving gently downward into the resolution
limit.  M~32-in-Virgo shows a nearly perfect power law with only
a small bend at the equivalent nearby radius of 70\asec.  Thus M~32 does not
fit Eq. (2) very well either, because of multiple breaks that yield
different values of $r_b$ depending on what portion of the profile is fitted.
M~32 shows that values of $r_b$ in power-law galaxies are not robust
and that similar breaks at small radii could exist in other distant power-law galaxies,
even those that apparently show clean power laws at the present resolution.

Because the fitted values of $r_b$ in power-law galaxies are less robust than
those for core profiles, which reflect real features, we regard them as less
fundamental. As explained below, we treat the fitted values of $r_b$
differently in analyzing the two types of galaxy. 

\bigskip\bigskip
\centerline{3. GALAXY SAMPLE AND DATABASE}
\bigskip

The database used in this paper is contained in Tables 1, 2, and 3.  A brief
overview is given here, and additional details are provided in the table
notes.  The heart of the sample consists of 42 normal ellipticals and bulges
taken from Paper I (NGC~4150, NGC~4826, and NGC~5322 were excluded due to
strong nuclear dust).  To these were added images of 14 E's and bulges from
the WFPC1 GTO programs (some unpublished).  Five more normal E's, mostly Virgo
galaxies from Jaffe {\it et al.}, were located in the HST public archive as of
June 1993, for a total of 61 galaxies.  The original GO/GTO program and
references to published {\it HST} profiles are listed in Table 1. All images
were taken using the Planetary Camera in Cycles 1 and 2 and consequently
suffer spherical aberration.  They were observed through filter F555W, which
approximates the V band, and usually have a peak signal of $\ge10^4$ photons
in the central pixel.  All images were processed as described in Paper I and
deconvolved with the same Lucy-Richardson procedure used there.

Power-law galaxies with identified nuclei are divided into two types:
``moderately" and ``severely" nucleated, indicated in Table 1 by ``$+$" and
``$++$".  M~31-in-Virgo is adopted as the dividing line between the two types
(cf. Fig. 2 here and Fig. 14 of Paper I).  In severely nucleated galaxies
and in galaxies with AGNs, fits to the nuker law ignore the innermost
pixels affected by the nuclear light.

Table 1 presents observed quantities such as Hubble type, distance, magnitude,
color, and nuker-law parameters from Paper II.  A few galaxies not
treated in Paper II have been similarly fit and the results are given here.
M~31 and M~32 appear twice, as seen nearby and near Virgo (labeled with a
``V").  For core-type profiles, we accept the nuker-law fits as given for
$\theta_b$ and $\mu_b$.\footnote{$^7$}{The quantity $\theta_b$ is the break
radius in arcsec, while $\mu_b$ is break surface brightness $I_b$ expressed as
V mag arcsec$^{-2}$.}  For power-law galaxies, no core is resolved, and we use
the separate upper limits on core size and surface brightness provided by
Paper I. These limits (for power laws only) are called $\theta_b^{lim}$ and
$\mu_b^{lim}$ in Table 1.  For a few power-law galaxies not contained
in Paper I, these limits were obtained from a visual estimate of the steepness
of the innermost part of the profile.

The distance to each galaxy (in \kms) has been estimated using a variety of
methods as summarized in the notes to Table 1, and the adopted value and its
conversion to Mpc (based on $H_0$ = 80 \kms Mpc$^{-1}$) are given there.
These distances are used to convert the apparent quantities in Table 1 to
absolute quantities in Table 2.  B-band magnitudes are converted to the V band
to be consistent with the {\it HST} profiles.  Data taken from the literature 
include central velocity dispersion,
$\sigma_0$, an inner velocity dispersion gradient defined as $R_{\sigma} \equiv
\sigma_0 / \sigma(10\asec)$, dimensionless rotation parameter $(v/\sigma)_*$,
isophote shape $a_4/a$, global (effective) radius $r_e$, and global surface
brightness $\mu_e$, defined as the mean surface brightness {\it within} $r_e$.
Details and sources are given in the notes.

Table 3 presents several derived quantities based on spherical, isotropic
dynamical models fitted to the nuker-law light profile.  
The mass-to-light ratio of each model has
been determined by normalizing to $\sigma_0$ from Table 1, assuming constant
$M/L$ with radius and equating $\sigma_0$ to the light-weighted rms
line-of-sight dispersion in a centered 2\asec\ by 2\asec\ aperture 
(corrections for 1\asec\ FWHM seeing are at most a few percent
and are not included).  Mass-related quantities are blank if $\sigma_0$
is not available.  Quantities tabulated at $0\s.1$
include the luminosity density, peak Maxwellian
phase-space density, two-body relaxation time, and predicted 
projected velocity dispersion.
Total luminosity and mass within a sphere of the same radius are also given.
Comparison to the non-parametric densities in Paper III indicates that
nuker-law fitted luminosity densities are 10\% too low on average but
otherwise show little scatter for non- and moderately nucleated galaxies
(severe nuclei were ignored in fitting nuker laws, and as a result nuker-law
densities in these galaxies are about a factor of $2$ lower than
the non-parametric inversions).  Several quantities
are repeated for $r = 10$ pc, but for many galaxies this is well inside the
resolution limit of $0\s.1$ and requires an inward extrapolation of the
nuker-law fit.  

An impression of the division into core and power-law galaxies is provided by
Fig. 3, which plots inner power-law slope $\gamma$ versus observed break
radius $\theta_b$ (or $\theta_b^{lim}$ for power-law galaxies) in arcsec.
Profiles with $\theta_b \ge 0\s.16$ (log $\theta_b \ge -0.8$)
are reasonably well resolved by {\it HST}.  They divide into two
groups, those with $\gamma \le 0.25$ (cores) and those with $\gamma > 0.5$
(power laws) --- the valley in between is empty.  This is the division 
that motivated the
two profile types in Paper I, later analyzed statistically in Paper III.

The rectangular box in Fig. 3 encloses galaxies that we are fairly sure
contain real cores.  Galaxies above the box are definitely power laws at
current resolution.  Galaxies to the left of the box are classed as power laws
although some contain a hint of an incipient core.  The effect of limited
resolution is apparent for M~31 and M~32; both galaxies are plotted
twice, as seen nearby and at Virgo.  The plotted positions differ appreciably,
reflecting features of their inner profiles that cannot be probed in more
distant galaxies.

Galaxies within the box in Fig. 3 comprise the ``Core" sample
used in the following section.  All others are classed as power laws.  

\bigskip\bigskip
\centerline{4. CENTRAL PARAMETER RELATIONS}
\bigskip

The data in Tables 1 and 2 are used to plot new central parameter diagrams
like those of Lauer (1983, 1985a) and Kormendy (1985, 1987a,b).  We begin with
plots versus absolute magnitude in Figs. 4a,b,c,d.  The symbols have the
following meanings:

\item{(1)} Core galaxies are plotted with filled circles ($\bullet$) using 
values of $r_b$ and $\mu_b$ from Table 2.

\item{(2)} Power laws are plotted with open circles ($\circ$) using the limits
$r_b^{lim}$ and $\mu_b^{lim}$ from Table 2.

\item{(3)} M~31 and M~32 are plotted twice, as seen at their actual
distance (asterisks) and in Virgo (end of vector).  The length and
direction of these vectors illustrate the possible effect of changing
resolution on other power-law galaxies.  Their direction is opposite to the
limit flags that are attached to all power-law galaxies.

\item{(4)} Special objects: The S0 galaxy NGC~524 is 
the only core profile that is found within a bulge (all others are in
ellipticals).  NGC~524 is roughly face-on and shows flocculent
dusty disk arms (Paper I) and a blue center (Kormendy, private communication);
it is plotted with a small square.  Fornax A (NGC~1316) is a
probable recent merger remnant (Schweizer 1980) with a peculiar
morphology (RC3).  It has an abnormally small core for a galaxy of
its luminosity (Kormendy 1987b).  NGC~4486B shows a double nucleus like M~31's in WFPC2
images (Lauer \et 1996) but continues to have a clearly defined core.

The new plots show the same broad trends versus galaxy luminosity that were
seen in ground-based data (Kormendy \& McClure 1993).  Core galaxies 
are luminous objects that extend down to $M_V=-20.5$.  All normal
ellipticals brighter than $M_V= -22$ show cores, with cores of brighter galaxies
being larger and lower in surface brightness and density.  The new central
parameters of core galaxies correlate well with previous values measured from
the ground (Kormendy \et 1994).  The parameter relations for core galaxies are
fairly narrow; for example, the rms scatter in $r_b$ versus $M_V$ about the
best-fitting line is only 0.25 dex (Fornax A omitted).

Ferrarese \et (1994) have questioned whether the trends in core properties
versus absolute magnitude are an artifact created by adding brightest cluster
galaxies (BCGs) to smaller, trendless galaxies.
They argue that, aside from M~87, all cores in their Virgo sample are of
similar size, and trends appear only when M~87 is added.  Although M~87 does
not strictly qualify as a BCG (that distinction in Virgo is held by NGC~4472), it
does share certain properties with BCGs such as high luminosity and central
location within a subcluster.

From our larger sample, it seems clear that trends in core properties versus
$M_V$ are real and are not an artifact of adding BCGs.  The present sample could be
truncated at $M_V = -22.2$ to eliminate {\it all}\ BCGs (including those in
small groups as well as Abell clusters), yet trends among the 11 remaining
core galaxies between $M_V = -20.5$ and $-22$ would still be present. In all
plots, core properties of BCG galaxies appear to be a normal extension of the
cores in smaller core ellipticals.

Power-law galaxies in Fig. 4 are low-to-intermediate luminosity systems that extend in
luminosity up to $M_V = -22$.  They overlap with core galaxies at intermediate
magnitudes in the range $-20.5 > M_V > -22$.  Despite an increase in angular
resolution by a factor of 10 with {\it HST}, we have generally
failed to find cores in these objects, and thus their distribution
in Fig. 4a is rather flat, reflecting the constant {\it HST} 
resolution limit of $\sim 0.1$ arcsec.
For systems fainter than $M_V \approx -19$,
this limit is uninteresting since it equals or
exceeds predictions based on extrapolation from core galaxies.  However, at
intermediate magnitudes in the range $M_V = -20.5$ to $-22$, power-law and
core galaxies coexist, and it is clear that the {\it scatter} in break radius is
real and large.  Core/power-law pairs that illustrate extremes of $r_b$ 
at fixed luminosity include
NGC~3379 and NGC~1023, whose break radii differ by more than a factor of 40
while their absolute magnitudes differ by less than 0.5 mag, and NGC~4168
and NGC~4594, for which the ratio of break radii is over 100  even though their
absolute magnitudes are almost identical.  
This is not a resolution effect
wherein cores are detected in nearby galaxies but not in distant ones.  Figure
5 plots break radius versus distance and shows that most of the sample,
containing both small and large cores, resides in a narrow range of distance
near that of Virgo.  More distant galaxies are actually {\it more} likely to show
cores because their cores are intrinsically larger.

The 
large scatter in break radii near $M_V = -20.5$ to $-22$ 
might at first sight be
taken as a manifestation of the two-dimensional, planar
distribution of the global structural parameters of hot galaxies, i.e.,
the {\it fundamental plane} (Dressler \et
1987; Djorgovski \& Davis 1987; Faber \et 1987).  Two-coordinate projections
of this two-dimensional distribution commonly exhibit scatter
depending on whether they show the plane edge-on or face-on.
The basic coordinates for the global plane (see Section 6) are
$r_e$, $\mu_e$, and $\sigma_0$, from which $L_V$ can be derived
as $L_V = 2 \pi \mu_e r_e^2$.  A plot of radius versus magnitude is thus
a projection of the fundamental plane, and 
scatter might be expected in $r_b$ versus $M_V$ that is comparable to that
seen in $r_e$ versus $M_V$, provided  $r_b$ and   $r_e$ are well
correlated. 

This hypothesis is tested by substituting $r_e$ 
for $r_b$ in Fig. 4d.  The scatter there 
proves to be small, demonstrating that
the combination of radius versus $L$ shows the {\it global} plane rather close
to edge-on.  The much larger scatter of Fig. 4a 
therefore suggests a {\it real
decoupling} of central properties from global ones, 
as emphasized by Lauer (1985a).
In Section 5 we examine this scatter in more detail and 
show that it correlates with global
rotation and isophote shape, in the sense that
power-law galaxies (which have small $r_b$)
are disky and rotate
rapidly, while cores (which have large $r_b$) are boxy and rotate slowly.  

Bulges are distributed in Fig. 4 like ellipticals of small-to-intermediate
size.  None (except for M~31-nearby) shows a core.  The resemblance of bulges
to small and intermediate ellipticals is not surprising since the two classes
of galaxy share several traits, including similar global size, high rotation,
flattening by rotation rather than anisotropy, and disky subsystems (Bender,
Burstein \& Faber 1992).

Before drawing further conclusions from Fig. 4, we consider whether
the trends shown there are affected by the
particular sample of galaxies chosen.  The present sample is
a mixture taken from different authors, but we have been careful
to retain only objects that are {\it morphologically
normal} and {\it free of dust}.  Our own sample from
Paper I (comprising 42 out of the 61 total objects in this paper) was 
specifically chosen 
to probe the full range of parameters covered by the ground-based central
parameter relations (Lauer 1985a; Kormendy \& McClure 1993).
We strove hard to sample the widest possible magnitude
range and, at intermediate magnitudes, to sample galaxies with both
large and small apparent cores.  Thus, it is possible that
the present sample somewhat exaggerates
the total spread in break radii at middle magnitudes.

Another point is that 
most objects studied here had previous
ground-based
data, and thus some prior clue as to core size.  Since
ground data typically agree well with {\it HST} data 
(especially for large galaxies, Kormendy \et 1994), 
the present sample does not provide a truly
fresh look at galaxy centers.  A sample to do this
with completely new galaxies has been
observed in Cycle 5 and is now being analyzed. 
What the present sample does is fairly probe galaxies that
had previously been examined from the ground.  

Are the claimed 
correlations robust for core galaxies specifically?  Although
Fornax A has been included in the diagrams for interest, it 
is strongly peculiar and its center is 
contaminated by dust (Shaya \et 1996).  It does not
qualify for our sample of normal, massive E's, and its
high residuals should not count against the correlations.
Six more core ellipticals with ground-based
data could also be added 
to bolster the {\it HST} data; these new galaxies agree well
with the trends here (Kormendy, unpublished).  Thus we feel that 
evidence strongly favors the core correlations found in Fig. 4; however, 
full confirmation will require the completely independent sample of galaxies
from Cycle 5 that we are now analyzing.

Finally, the number of galaxies at intermediate magnitudes
is still small.  We will argue in the next section that the
spread of properties in this magnitude range is correlated with global
boxiness and diskiness, and will draw from this some
significant conclusions about hot galaxy formation.  Clearly,
these conclusions will need to be checked by gathering a larger and
more representative sample of galaxies at these 
magnitudes.  Again, the Cycle 5 sample was selected to do this.

The referee has asked whether bright power-law galaxies might 
in fact all be S0's (or bulges).  There are 7 power-law galaxies in the magnitude
range $-$20.5 to $-$22: NGC~596 (E$^+$4:),  NGC~1172 (E$^+$2:),  NGC~1700 (E4),
 NGC~3115 (S0$^-$), NGC~4594 (Sa), NGC~4621 (E5), and NGC~4697 (E6)
(Hubble types from the RC3).  Two of these
are actual disk galaxies (NGC~3115, NGC~4594), two more are highly flattened
E's (NGC~4621, NGC~4697), and two more have S0-like outer envelopes
(NGC~596, NGC~1172).  That leaves only NGC~1700, which is also
fairly flattened.  We will show in the next section that there is
a good correlation between power-law galaxies and galaxies that
are rapidly rotating with disky isophotes.  Thus it is 
possible that all bright E-type power-laws are in fact 
S0's masquerading as ellipticals.

It is interesting to speculate where the bright power-law galaxies will
move within Fig. 4 as resolution improves.  At present these
galaxies fall below the core sequence
by $\times$3-10 in core size.  However, new WFPC2 observations
have increased this distance for a number of them (Lauer \et 
1997).  It is thus possible that bright power-law galaxies may
ultimately emerge as a separate population 
rather than simply tracing the lower edge
of a large dispersion in core size at these luminosities.

M~31 and M~32 again show the effect of changing spatial resolution.
M~32-in-Virgo lies near the Virgo dwarf E's in all of its parameters,
suggesting that the downward trend in surface brightness for faint galaxies in
Fig. 4c is mainly an artifact of resolution.  M~31-in-Virgo is
indistinguishable from other bulges of similar magnitude.  Its shift in the
diagrams under distance change is not as large as that of M~32 because its
profile is not as steep as M~32's between $0\s.1$ and $2\s.4$ (Fig. 2).

NGC~4486B is the lone core galaxy near $M_V = -17.5$ mag.  Its angular
size lies near the lower boundary of established cores in Fig. 3, but the
presence of a core has been confirmed in WFPC2 images (Lauer \et 1996).
NGC~4486B's low luminosity, compact profile, high line-strength, and close
proximity to M~87 suggest that it might be tidally stripped by its
larger neighbor (Faber 1973).  Its core parameters would be consistent 
if it once resembled the small-core galaxy
NGC~3608 and then lost $\sim$90\% of its outer luminous envelope.  
Whether a core
could actually survive such extensive  stripping and whether a diffuse giant like
M~87 could strip a high-density object like NGC~4486B are open questions (see
Section 7.3.2).

Figure 6 plots luminosity density, mass density, and phase-space density at a
limiting radius of $0\s.1$ (from Table 3) as a function of absolute
magnitude.  An impressive feature of Fig. 6 is the large range 
in density from small to
large galaxies, almost 10$^6$ in all three panels; half of this range is
spanned by cores alone.  Part of this spread is due to the
fact that Fig. 6 mixes objects at different distances.   However, 
densities at the fixed physical scale of 10 pc still show a range of 1000
(column 13, Table 3; see also Fig. 13, Paper I; note that densities
in distant galaxies at 10 pc require inward extrapolation).
The large range of densities near the centers of hot
galaxies has been remarked on before (Kormendy 1984, Lauer 1985a, Carlberg
1986), but {\it HST}'s higher resolution has pushed up densities in power-law
galaxies by another two orders of magnitude.  As discussed in Section 7, 
the large density contrast
between small and large hot galaxies is an important clue to hot galaxy
formation. 

We conclude this section by presenting general scaling laws for core
properties versus galaxy luminosity and mass.  We restrict attention to core
galaxies
because their parameters are robust.  It is well known (see Section 6) that
cores are really a {\it two-dimensional} dynamical family (the fundamental
plane).  Nevertheless, it is often convenient to treat them as a one-parameter
family depending on luminosity or mass. This is possible because
(1) the fundamental plane is only partially filled, and projections
against any pair of coordinate axes have only limited extent; (2) typically,
these projections are elongated and can be approximated by 
one-dimensional scaling relations; and (3) core mass and luminosity are 
both moderately
well related to global mass and luminosity for core galaxies.  
This combination 
produces the tight correlations versus total luminosity 
in Figs. 4 and 6, which we now quantify.

The following is a set of self-consistent scaling relations versus $L_V$ and
galaxy mass, $M$.  Total luminosity has been related to mass by assuming $(M/L_V)
\propto L^{0.25}$ (Faber \et 1987).
The exponents in these relations are not least-square fits but have been
derived by a process of trial-and-error adjustment to maintain consistency
with standard structural formulae. 
The first three of these relations are independent fits to the data, while the
rest are derived from the structural formulae.
The relations involving $M_{core}$ assume that
the core is in dynamical equilibrium.

$$      r_b    \propto L_V^{1.15}                 \propto M^{0.92},                        \eqno{(3)}  $$
$$      I_b    \propto L_V^{-1.0}                 \propto M^{-0.8},                        \eqno{(4)}  $$
$$ \sigma_0    \propto L_V^{0.2}                  \propto M^{0.16},                        \eqno{(5)}  $$
$$ j_{core}    \sim  {I_b \over r_b}              \propto L_V^{-2.15}  \propto M^{-1.72},   \eqno{(6)}  $$
$$ L_{core}    \sim  {I_b r_b^2    }              \propto L_V^{1.3}    \propto M^{1.04},     \eqno{(7)}  $$
$$ M_{core}    \sim  {\sigma_0^2 r_b}               \propto L_V^{1.55}   \propto M^{1.24},    \eqno{(8)}  $$
$$ M_{core}/L_{core}  \sim  {\sigma_0^2 \over {I_b r_b}}   \propto L_V^{0.25}   \propto M^{0.2},    \eqno{(9)}  $$
$$ \rho_{core} \sim  {{\sigma_0^2 } \over r_b^2} \propto L_V^{-1.9}   \propto M^{-1.52}.    \eqno{(10)}  $$

\bigskip\bigskip
\centerline{5. SCATTER IN THE CENTRAL PARAMETER RELATIONS:}
\centerline{CORRELATIONS WITH GLOBAL ROTATION AND ISOPHOTE SHAPE}
\bigskip

An emerging suspicion of the last decade is that there
are actually {\it two} types of elliptical galaxies: luminous E's
with boxy isophotes that rotate slowly, and small E's 
with disky isophotes that rotate rapidly (Bender 1988; Nieto 1988;
Nieto \& Bender 1989; Bender \et 1989; Kormendy
\& Djorgovski 1989).  A 
formal division of the Hubble sequence for ellipticals into two classes has
been suggested based on these criteria (Kormendy \& Bender 1996).  We shall
refer to these two subtypes as {\it boxy} and {\it disky} respectively.

The scatter seen in central parameters
versus absolute magnitude in Figs. 4 and 6 appears to correlate with boxy/disky
subtype.  This correlation is illustrated in Fig. 7, which replots Fig. 4a
($r_b$ versus $M_V$) but now with symbols indicating rotation, $(v/\sigma)_*$,
in Fig. 7a, and isophote shape, $a_4/a$, in Fig. 7b.  It is seen that
power-law galaxies are
mainly rapidly rotating and disky, while cores are slowly rotating and boxy or
neutral.  {\it At intermediate magnitudes, $M_V = -20.5$ to $-22$, the
presence of a core is a better predictor of boxiness or slow rotation than is
absolute magnitude.}  The same correlation was also found by Nieto \et (1991a)
using ground-based data.  In our sample, the
correlation between disky galaxies and power laws appears strongest if the
disky threshold is set at $a_4/a \geq 0.4$.  All other isophote types (boxy,
neutral, and variable) are associated with cores.  ``Variable" galaxies
are those with strongly varying values of $a_4/a$ versus radius.  We use the
term {\it boxy} to include the boxy, neutral, and variable types.

Jaffe \et (1994) and Ferrarese \et (1994) also saw a link between
power-laws and disks based on their {\it HST} Virgo
sample.  They went further to suggest that essentially all power laws have
{\it inner} disks within 1\asec \ and that such disks seen edge-on are what
produce power-law profiles (Jaffe \et 1994).  We agree that power laws tend to
reside in galaxies with {\it globally} high rotation and global diskiness, 
but we do not see evidence for {\it inner}
disks in all or even most power laws.  Rather, we believe that 
the {\it hot} 
component is intrinsically different in cores and power
laws and that the two classes therefore 
would look different from any viewing angle.

This difference in interpretation is fundamental, since implications for
galaxy formation would be limited if profiles were a sensitive function
of viewing angle.   We believe that Jaffe \et were influenced by the fact
that most power-law objects in their sample happened to be highly
flattened, edge-on, late-type E/S0's with a higher-than-average incidence of
both inner and outer disks.  Our power-law galaxies as a group are less
flattened, less edge-on, less skewed to late Hubble types, and do not in
general show inner {\it or} outer disks.  
A brief discussion of the edge-on disk model for power laws
was presented in Paper I.  A more extensive search for inner disks was made
with the present larger sample, and the results are reported in Appendix A.

To summarize, it appears likely that disky and boxy ellipticals have different
kinds of central light profiles.  Since {\it global} properties are
implicated, whatever process established this connection was probably a major
event in the life of the galaxy.  We explore this link and its implications
for hot galaxy formation in Section 7.

\bigskip\bigskip
\centerline{6. THE CORE FUNDAMENTAL PLANE} 
\bigskip

So far we have focussed on the
relationship between the centers of galaxies and their global properties.  We
turn next to relationships among the central properties alone.  Plotting
galaxies in central $(\log r_b,\mu_b,\log\sigma_0)$--space allows us to look for a
fundamental plane (FP) analogous to the one found in global $(\log
r_e,\mu_e,\log\sigma_e)$--space (Dressler \et
1987; Djorgovski \& Davis 1987; Faber \et 1987).  If cores are in dynamical
equilibrium (highly likely), if they are supported by random motions (as
indicated by the observations), if velocity anisotropy does not vary too much
from galaxy to galaxy (unknown, but see below), and if core $M/L$ is a
well-behaved function of any two variables $\mu_b$, $r_b$, or $\sigma_0$ (true
of global $M/L$), then we expect cores to populate a thin surface in central 
$(\log r_b,\mu_b,\log\sigma_0)$--space 
(Faber \et 1987).  Lauer (1985a) demonstrated that cores of
well-resolved galaxies indeed populate a two-dimensional manifold, and Faber
\et (1987), using Lauer's data, derived a preliminary core fundamental plane
that was roughly parallel to the global FP.  We revisit the tilt and thickness
of this plane using the more accurate {\it HST} data.

Figure 8 shows projections of $(\log r_b,\mu_b,\log \sigma_0)$--space for the
present sample.  Cores are again filled circles, while power-law galaxies
(open circles) have been plotted for completeness' sake using their limiting
values (warning: the direction of the 
limit flags is schematic using M~31 and M~32 as a guide).
To seek a plane, we rotate
about an axis and search for the thinnest distribution of points.  Following
Faber \et (1987), we choose to rotate about the $\sigma_0$-axis because the
resulting combination of $I_b$ and $r_b$ is nearly invariant to small
resolution changes and small measurement errors.  The best rotation (based on
core galaxies only) is shown in Fig. 8d.  Within the errors, the tilt of
this FP is consistent with the global plane, 
$\sigma^{1.35} \sim r_{e} I_{e}^{0.84}$, found by Faber \et (1987); this is the
orientation plotted. The residual rms scatter about this plane, expressed as
an error in $\log r_b$, is 0.12 dex (for cores only).  This is 30\% larger
than the equivalent scatter about the global FP, which is 0.09 dex
(Lynden-Bell \et 1988). The larger scatter about the core plane may be related
to the presence of central BHs, which elevate $\sigma_0$ in some galaxies.
Five labeled objects with positive residuals in Fig. 8d are BH candidates
from Kormendy \& Richstone (1995) and Kormendy \et (1996b,c).

The locus of $I_b$ versus $r_b$ is also quite narrow for cores (Fig. 8c).
This occurs because the individual brightness profiles of core galaxies are
approximately tangent to a single line in the $(\log r_b, \log I_b)$ plane, as can be
seen in Fig. 1.  This tight correlation can be used to construct a second
distance indicator based on $r_b$ and $I_b$ alone.  The scatter about the
best-fitting line corresponds to 0.18 dex in log distance, which is two times
worse than the global FP.  However, the method does not require any
measurement of $\sigma$ and thus may sometimes be useful.

The existence of a core fundamental plane suggests that (1) cores are in
dynamical equilibrium supported by random motions; (2) that $r_b$ and $I_b$
are meaningful dynamical parameters describing the size and luminosity density
of the core; (3) that velocity anisotropy does not vary greatly among core
galaxies; (4) that the mass of any central BH does not strongly dominate the
core potential in most galaxies; and (5) that core mass-to-light ratio varies
smoothly over the fundamental plane.

The core fundamental plane is well defined even though (1) the profiles of
core galaxies are {\it not} analytic; (2) the profiles of different core
galaxies are not identical ($\alpha$ and $\gamma$ vary, Paper I); (3) the
velocity dispersion anisotropy may vary from galaxy to galaxy; (4) some or all
core galaxies may harbor massive BHs that distort both the photometric profile
and $\sigma_0$; and (5) cores represent only a tiny fraction of the total
luminosity of the galaxy.  Evidently whatever differences exist among core
galaxies are not so large as to erase the appearance of a two-parameter family
of self-gravitating cores that is fundamentally not too dissimilar from the
2-dimensional family of isothermal spheres.

\bigskip\bigskip
\centerline{7. CENTRAL PARAMETER RELATIONS AND HOT
GALAXY FORMATION}
\bigskip

The final sections of this paper discuss the central parameter relations 
in the context of galaxy formation.  We assume throughout
that hot galaxies form via hierarchical clustering and merging (hereafter 
HCM; see e.g., Toomre 1977, White \& Rees 1978,
and Blumenthal \et 1984).  Descriptions of HCM 
as applied to hot galaxies 
may be found in Schweizer (1986), Kormendy \& Sanders (1992),
de Zeeuw \& Franx (1991),  Barnes \& Hernquist
(1992), and Bender \et (1992). 
An important challenge to HCM is the
formation of boxy and disky galaxies, including the 
association found here with central cores and power laws.
A novel element that needs to be considered
is the  presence of massive BHs in the
centers of many or most hot galaxies, which could considerably 
alter the predictions of standard HCM  with no BHs.

We will suggest that the high-density cusps of power-law
galaxies are broadly consistent with the growing body of evidence 
that points to the importance of {\it gaseous dissipation}
in their formation.  More of a puzzle are 
the low-density cores of massive hot galaxies
--- they seem hard to form in the
first place and hard to maintain once formed.
The problem of cores leads us to consider an alternative
method for forming them based on central BHs.  This model is the
subject of Section 8.

\bigskip
\centerline{\it 7.1 Previous Work on Hot Centers}

Many authors have discussed the centers of hot galaxies in the context of
galaxy formation. Larson (1974a,b) computed gaseous collapse
models for ellipticals and noted that central star formation could continue
until very late, fueled by dregs of gas falling into the center.  He
conjectured that central star density might depend on a delicate balance
between the dissipation rate, global star formation efficiency, and mass loss
via supernova-driven winds.

Lauer (1983, 1985a) discussed the systematic properties of the centers of hot
galaxies in light of new data.  The 
high-density centers of small galaxies were consistent with Larson's gaseous
infall picture, but the same theory predicted greater gas retention, and
hence denser centers, in the deeper potential wells of bright galaxies,
contrary to observations.  An alternative scenario based on dissipationless
merging also ran into difficulties because N-body
simulations showed denser cores forming as galaxies successively merged
(Farouki \et 1983); this ``departure from homology'' has been confirmed
with more modern $N$-body simulations (Barnes 1992; Makino \& Ebisuzaki 1996).

Carlberg (1986, see also Ostriker 1980) used phase-space density arguments to constrain
the progenitors of hot galaxies.  He noted that the high phase-space densities
of small galaxies precluded their formation from {\it purely stellar} spiral disks
because the phase-space density of spiral disks is low and 
phase-space density in dissipationless mergers remains constant or declines 
(Barnes 1992).  However,
HCM naturally incorporates gaseous merging as well as stellar merging
(Schweizer 1986; Kormendy 1989; Kormendy \& Sanders 1992), either during the
main merger event (Negroponte \& White 1983; Barnes \& Hernquist 1991;
Hernquist \& Barnes 1991), during an earlier phase in which the disk
progenitors develop bulges (Barnes 1992; Hernquist 1993), or 
during later gaseous infall.  Thus it is no problem to create the 
high densities of small hot galaxies 
provided gaseous dissipation is present at some stage.

The above papers considered mainly 
equal-mass mergers, but small satellites can also be accreted by dynamical friction
(e.g., Tremaine 1976).  Kormendy (1984)
applied this concept to the capture of small, dense ellipticals by larger
ones (see also Balcells \&
Quinn 1990) and predicted the photometric and kinematic signatures of such events:
cores-within-cores, high surface-brightness centers, central velocity
dispersion dips, central counter-rotation, and isophote twists.  Many of these
anomalies have since been found (de Zeeuw \& Franx 1991;
Barnes \& Hernquist 1992), leading to the concept of (photometrically and/or
kinematically) {\it decoupled centers}.  Altogether, roughly a third of hot
galaxies show such anomalies (de Zeeuw \& Franx 1991), with core galaxies
showing them roughly twice as often as power laws (Nieto \et 1991b, Paper
I).  

Hernquist \& Barnes (1991) suggested an alternative way
to make decoupled centers involving gaseous accretion and subsequent star
formation in a cold inner disk.  Decoupled subsystems are often 
dynamically colder than
expected from purely stellar satellite accretion
(Bender \& Surma 1995,
Franx \& Illingworth 1988).  Their stars are also
stronger-lined than stars at the centers of
small satellites, suggesting that local nucleosynthetic
enrichment (and hence star formation) has taken place (Bender \& Surma, 1988). 

To summarize, the consensus exists that gaseous dissipation
plus {\it in situ} star formation are the key factors
responsible for the high central densities of small hot galaxies.
Both cores and power-laws 
probably also contain both stellar and gaseous material
captured in accretions and mergers.  Less clear
is why central properties scale at all with galaxy mass and in
particular why the centers of massive hot galaxies are so diffuse.

\bigskip
\centerline{\it 7.2 Power Laws in Disky Galaxies}

We turn now to the relation between central and global properties.  The basic question
is whether the core/power-law distinction found for the inner parts is
consistent with theories for forming the outer parts.

The situation seems 
clearest for disky galaxies.
The formation of disky 
hot galaxies, though not fully understood within HCM
(Kormendy \& Bender 1996), probably stems from the presence of
significant quantities of {\it gas} during the latest merger(s).  Several
authors have noted that the high rotation, disky isophotes, and lack of
minor-axis rotation of disky ellipticals imply global gaseous dissipation 
(Kormendy 1989; Nieto \et 1991a; Bender \et
1992).  Recent simulations with gas (Barnes 1996,
Barnes \& Hernquist 1996) show that just a few percent
of the mass in gas is sufficient to destroy box orbits and impart high global rotation;
the same gas can make disky isophotes if it forms stars.

The steep central power laws of disky galaxies are 
plausible by-products of such gaseous mergers.  
Numerical models of gas-rich mergers (Barnes
\& Hernquist 1991, Hernquist \& Barnes 1991, Mihos \& Hernquist 1994) have
shown that angular-momentum transfer and dissipation can swiftly carry much of
the gas in merging galaxies to the center of the remnant.  Strong evidence for
efficient gaseous infall is seen in nearby gas-rich merger remnants, which possess
massive central clouds of gas (Sanders \et 1988, Scoville \et 1991) that may be
fueling central starbursts (Bushouse 1987; Kennicutt \et 1987; Leech \et 1989;
Condon \et 1991; Solomon, Downes \& Radford 1992; Kormendy \& Sanders 1992).
A major uncertainty is exactly where and how the stars form, and hence 
the shape and density of the resulting stellar profile; if anything,
current models of starburst cusps are {\it too} dense and compact (Mihos \&
Hernquist 1994).  This problem aside, the high-density cusps of power-law
galaxies seem broadly consistent with their formation in dissipative,
gas-rich mergers\footnote{$^{8}$}{
It may be significant that 6 out of 7 power-law galaxies in our sample with
$-20.5 > M_V > -22$ are {\it field} objects, whereas 6 out of
9 core galaxies in the same range are in {\it clusters}.  Kauffmann (1996),
using Press-Schechter theory, has proposed that the clustering history of
intermediate-magnitude ellipticals depends on environment --- those in clusters
are old, while many in the field are formed from recent mergers of gas-rich
spirals.  A preponderance of power-laws in field galaxies would be
consistent with their formation in recent gas-rich mergers.  In fact, two of the
6 field power-law galaxies are known 
merger remnants (NGC~596 and NGC~1700; Schweizer \et 1990).}.

\bigskip
\centerline{\it 7.3 Cores and Boxy Hot Galaxies}

The above discussion supports the notion that 
disky galaxies and their central power-laws were formed 
together in gas-rich mergers.
Analogous  arguments suggest that
boxy galaxies formed in gas-poor mergers (Binney \& Petrou 1985;
Bender \& M\"ollenhoff 1987; Nieto 1988; Nieto \& Bender
1989; Nieto \et 1991a; Bender \et 1992).  The distinctive shape, slow
rotation, anisotropy, and minor-axis rotation of boxy galaxies are
consistent with a large population of stars 
moving on box orbits in a triaxial potential created 
during a dissipationless merger (Barnes 1988, 1992).

We therefore ask: are the core profiles of boxy galaxies 
simply the natural by-product of dissipationless stellar 
merging?  To address this, the merging history of boxy galaxies
can be simplified into two parts: an
early phase in which centers originally formed, and a later phase involving the accretion of
small satellite companions.  Do core profiles
form early, and do they survive later accretion?  

\bigskip
\centerline{\eightit 7.3.1 Early Formation}

The early formation of boxy, core galaxies is murky 
because their progenitors are poorly known\footnote{$^{9}$}{It is
clear, however, that bright ellipticals were {\it not} formed by simply
merging {\it today's} faint ellipticals.  This is precluded by their
very different stellar populations (Bender \et 1992) --- each type must have had
its own progenitors.}.  
Existing $N$-body simulations of
dissipationless equal-mass mergers do not develop
cores --- rather, pre-existing cores tend to shrink slightly due to
non-homology, and the central density increases at each level of merging
(Farouki \et 1983; Barnes 1992; Makino \& Ebisuzaki 1996). Thus it appears 
that cores in luminous galaxies do not arise spontaneously in equal-mass
merging, although the resolution of present N-body experiments is limited.

The problem may be worse with {\it un}equal-mass mergers, in which a
smaller, denser component could sink to
the middle, perpetuating a high-density center.  This situation is
discussed further under satellite accretion.  On the other hand, progenitors
of boxy galaxies may differ from today's hot galaxies and may not obey 
the same inverse correlation between mass and density.

An entirely different way to generate low-density cores via stellar merging
is to start with
pure spiral disks. However, conventional
density fluctuation spectra do not form spirals 
in the overdense environments that
give rise to elliptical galaxies (Blumenthal \et 1984, Bardeen
\et 1986).  Yet a third way to make diffuse cores is via mass loss in
stellar or AGN-driven winds, but the deeper potentials of luminous core
galaxies should retain more gas, not less (Larson 1974b).  A final possibility
is to whip phase-space vacuum into centers during merging, for example via
mergers of multiple subclumps (Weil \& Hernquist 1996); however, it appears
that the merging
of the subclumps must be nearly simultaneous, which would be difficult
to orchestrate for every core galaxy.

\bigskip
\centerline{\eightit 7.3.2 Late Satellite Accretion}

The possible difficulty of forming cores may be matched or superseded by 
the even greater problem
of maintaining them against satellite infall.
In any merging hierarchy, the more luminous galaxies cannibalize the less
luminous ones.  It is plausible that the central region of the smaller galaxy
will survive intact so long as its radius is smaller than the tidal radius
imposed by the larger galaxy; this in turn implies that the
regions of the smaller galaxy that are denser than the core of the large
galaxy should survive. 

In fact, the centers of today's satellite galaxies are {\it much}
denser than the centers of core galaxies (see Fig. 13, Paper I); typical
power-law galaxies in the range $M_V = -17.5$ to $-22$ are 100 to 1000 times
denser at 10 pc than Abell brightest cluster galaxies (BCGs), 
and are 3 to 30 times denser at 100 pc, the
inner boundary where accreted satellites can be detected by {\it HST}. 
The tidal argument therefore suggests that dense satellites
should survive infall, filling in low-density cores of bright 
galaxies as proposed by Kormendy (1984).  Yet every bright galaxy in our sample
(except Fornax A) has a low-density core.

To quantify this paradox, we estimate a typical satellite accretion rate
for Abell BCGs. From counts of nearby companions and other
data, Lauer (1988) deduced an accretion rate for BCGs of 0.2$L^*$ per Gyr, in
close agreement with a theoretical estimate by Merritt (1985).
There are 7 Abell BCGs in the current {\it HST} sample, all of which have
large, low-density cores.\footnote{$^{10}$} {They are NGC~2832, NGC~4889, NGC~6166,
NGC~7768, Abell 1020, Abell 1861, and Abell 2052.} If 
satellite profiles are preserved during infall, accreted satellites over a particular
magnitude range will be detectable.  Small satellites have too little light,
while large ones have profiles that are too similar to the BCG to make a
difference. The profiles in Fig. 1 imply \footnote{$^{11}$}
{Our criterion is that the net profile after infall be one magnitude brighter
than presently observed at the inner resolution limit.  This is sufficient
either to erase a core in marginally resolved galaxies or create a tell-tale
inner upturn in well-resolved cores.}  that satellites with $M_V$ between
$-19.0$ and $-22.0$ would be detectable in all 7 BCGs, and that those
between $-17.5$ and $-22.0$ would be detectable in all but Abell 2052.  If
BCGs have been accreting for 5 Gyr at the rate estimated by Lauer (1988), this
translates to 2 detectable accreted satellites per BCG, or
13 total accretions in 7 galaxies.

This estimate is conservative --- gaseous accretion has been neglected, and the
current accretion rate by BCGs is probably lower than average owing to the
rise in cluster velocity dispersions with time.  Restriction to Abell
BCGs has excluded such near-BCGs as NGC~4874 in Coma (which was probably once
the BCG of its subgroup), Virgo's BCG, and BCGs of smaller groups like
Pegasus, Fornax, and Eridanus (where accretion is probably faster owing
to smaller velocity dispersions).  Including such objects would double the
number of primaries to 14 and raise the number of expected accretions to 
26.  However, no filled-in cores are seen in any of these BCGs.

So far we have assumed that the inner portions of accreted satellites survive
while sinking to the centers of their primaries, based on 
the tidal disruption argument.  
This argument has been criticized by Weinberg (1994, 1997),
who stresses that the time-dependent tidal force from the host galaxy can
do work on resonant stars in a satellite galaxy even when the satellite is
much denser than the host.  Using semi-analytic perturbation theory
and King-model profiles, Weinberg
concludes that, if satellite and primary obey the global fundamental
scaling law of Eq. (10), 
the satellite will be disrupted during its orbital decay if its mass
exceeds $10^{-3}$--$10^{-2}$ of the primary mass.  If this were true,
low-density cores would remain unaffected by late accretion because
of satellite disruption.

The most relevant N-body simulation so far of satellite survival 
is a merger of two pure ellipticals with mass ratio 10:1 
by Balcells \& Quinn (1990).  The
density scaling of the small galaxy relative to the larger one approximately
follows Eq. (10).  At the end of the merger, 
the pre-existing core of the primary is filled in by an amount
that would be detectable by {\it HST}.  
Further N-body models are in progress to check 
Weinberg's analytic results (Dubinski 1997) and to
simulate the dense central power-laws of real satellite
galaxies (Minske \& Richstone 1997).  Realistic modeling
of these dense centers may prove crucial.

\medskip
To summarize this section, the link between the centers of hot
galaxies and their outer parts must be accounted for in the
HCM picture.  Many properties of both disky and 
boxy galaxies are naturally explained by appealing to a difference in 
the amount of gas present during the most recent
merger(s).  Disky galaxies, including their high central
densities, suggest final mergers that were
{\it gas-rich}.  Analogous arguments concerning boxy galaxies are less clear: the
global kinematics of these galaxies suggest final mergers that were {\it gas-poor}, 
but forming and preserving cores in such models
may be difficult.  An enlargement of the HCM model for core formation 
that includes BHs is considered 
in the next section.

\bigskip\bigskip
\centerline{8. CORE CREATION BY MASSIVE CENTRAL BLACK HOLES}
\bigskip

High-resolution kinematic observations of galaxy centers strongly indicate
that massive BHs are normal constituents of the centers of hot galaxies
(Kormendy \& Richstone 1995). Three new BH candidates have been
discovered (Ferrarese \et 1996 [NGC~4261]; Kormendy \et 1997 [NGC~4486B];
Bower \et 1997 [NGC~4374]), and
the case for 5 more has been strengthened (Harms \et 1994 [M~87], Kormendy \et
1996b [NGC~3115], Kormendy \et 1996c
[NGC~4594], Gebhardt
\et 1997 [NGC~3377], and van der Marel \et 1997 [M~32].  BHs may play a key role in
determining the central structure of galaxies, and no discussion of the
central structure expected in HCM models would be complete without examining their
influence.

If both BHs and mergers are common among hot galaxies, two galaxies
with pre-existing BHs will frequently merge. The BHs will spiral towards the
center of the merger remnant, heating and perhaps ejecting the stars. This
process may form the observed core in the merger remnant (Begelman \et 1980;
Ebisuzaki \et 1991; Makino \& Ebisuzaki 1996; Quinlan 1997, Quinlan \& Hernquist 1997).

In what follows we assume that every hot galaxy contains a BH with average
mass
$$       M_{\bullet} = 0.002~ M_{gal},    \eqno{(12)}$$
where $M_{gal}$ is the mass of stars in the spheroid.  The adopted coefficient
0.002 is a mean of estimates based on the energetics of AGNs and direct mass
estimates of local BHs (see Appendix B).  The assumption of proportionality in
Eq. (12) is motivated by current data on local BHs (Kormendy \& Richstone
1995; Kormendy \et 1997), 
although the measurements show a scatter of at least an order of
magnitude.

The evolution of a pair of BHs in a merger remnant was first examined by
Begelman \et (1980). A recent comprehensive analysis is provided by Quinlan
(1997).  The two BHs are carried toward the center of the remnant by the
general inward motion of the dense central parts during the merger but
will not be exactly at the center at the end of the main 
merger phase.  Subsequent
migration of the BHs towards the center occurs on a slower timescale via
dynamical friction from the background sea of stars. As the BH orbits decay,
they form a bound binary BH whose semi-major axis $a$ continues to shrink
through dynamical friction. As the binary becomes more tightly bound,
dynamical friction becomes less effective, and the characteristic decay time
$|d\log a/dt|^{-1}$ increases. Finally the binary orbit shrinks to the point
that gravitational radiation or gas accretion dominates the decay, and rapid
coalescence ensues.  

Decaying BHs lose most of their energy by heating the surrounding stars.  The
consequent puffing up of the galaxy  was first examined by Ebisuzaki
\et (1991).  Based on rough analytic arguments and $N$-body models, they 
proposed that a merger of two galaxies with BHs would create a 
low-density core even if
none previously existed.  They argued that the mass of this core is
approximately equal to the sum of the BH masses $M_{\bullet} = m_1 + m_2$.  If
all galaxies start with the same ratio of BH mass to galaxy mass, this ratio
would be unchanged by later merging, and thus the ratios $r_b/r_{e}$ and
$M_{\bullet}/M_{gal}$ would remain constant.  This was later seen in
hierarchical merging N-body experiments with BHs (Makino \& Ebisuzaki 1996)
and also agrees approximately with observations (see below).

Quinlan and Hernquist have reexamined the evolution of binary BHs using scattering
experiments and $N$-body models (Quinlan 1997, Quinlan \& Hernquist 1997).
The following discussion
is based on their results, which treat the infall of equal-mass pairs of BHs
ranging in individual mass from 0.00125 to $0.04M_{gal}$; the total BH mass
$M_\bullet=m_1+m_2$ ranges from 0.0025 to $0.08M_{gal}$.  Quinlan
has kindly provided details of these models, which allow us to estimate the
ratio of BH mass to core mass, a key quantity needed to compare to
observations.

The  models start with a spherical galaxy whose density profile follows 
either a Hernquist law,
$$ \rho(r) \propto {1\over {r \left(1 + {r / r_s} \right)^3}},
                                                            \eqno{(13)}
$$
or a modified Hernquist law with steeper slope in the inner parts:
$$ \rho(r) \propto {1\over {r^{1.5} \left(1 + {r / r_s} \right)^{2.5}}}.
                                                            \eqno{(14)}
$$
The BHs are started on circular orbits at the half-mass radius $r_{e}$ of the
galaxy. As the BH orbits shrink by dynamical friction, the stellar profiles
develop cores.  Final break radii $r_b$ were
measured (by us) by locating the maximum of the logarithmic curvature of the
projected mass surface density, which is equivalent to the definition in the
nuker law.  We define the indicative core mass as
$$ M_{core} \equiv \pi r^2_b \Sigma(r_b),
                                                            \eqno{(15)}
$$
where $\Sigma(r_b)$ is the projected surface density at $r_b$.  The
ratios $M_{core}/M_{\bullet}$, $M_{core}/M_{gal}$, and $r_b/r_{e}$ were
tabulated for every model.  $M_{gal}$ is the stellar galaxy mass
(dark matter is ignored). 

The resultant 
core mass is approximately proportional to the BH mass but depends
somewhat on the mean slope of the original mass profile over the region
covered by the new core. For $M_{\bullet}/M_{gal}$ near 0.002 (Eq. 12) 
we find 
$$ M_{core} = (3.5 - 6.4) M_{\bullet},
                                                            \eqno{(16)}
$$
which translates to
$$ M_{core} = (0.007 - 0.012) M_{gal}
                                                            \eqno{(17)}
$$
if $M_{\bullet}/M_{gal} = 0.002$.  The range in parentheses reflects the two
models in Eqs. (13) and (14). The scaling relation for 
break radii analogous to Eq. (17) is found to be 

$$ r_{b} = (0.02 - 0.06) r_{e}.
                                                            \eqno{(18)}
$$

These results imply that the orbital decay of a BH creates an indicative core
mass that is 3-6
times the BH mass.  This is larger than the core mass $M_{core} \sim M_{\bullet}$ 
estimated
by Ebisuzaki \et (1991), and larger than the ejected mass 
$M_{ej} \ltsim2 M_{\bullet}$ found by Quinlan \& Hernquist.  Our explanation
is that the ejection of a given mass can create the impression of a more
massive core simply due to the precise definition of $r_{core}$ as the
point of maximum logarithmic curvature in the profile --- the exact
location of  $r_{core}$ depends sensitively on how it is defined.  The larger indicative
core mass in Eq. (16) comes closer to matching the observed indicative core mass for
M~87, which is $\sim10M_{\bullet}$.\footnote{$^{12}$}{Based on the core parameters of M~87 in
Table 2, the fitted nuker-law global $M/L_V$ value for M~87 of 10.2, and
$M_{\bullet}$ = 3 $\times 10^9$ M$_{\solar}$ (Harms \et 1994, as scaled
by Kormendy \& Richstone 1995).}

The theoretical predictions of Eqs. (17) and (18) are compared to observed core
luminosities and break radii in Figs. 9a and 9b.  Indicative core luminosity
for observed galaxies is defined (analogously to $M_{core}$) as $L_{core}
\equiv \pi r_b^2 I_b$ and is computed from the core parameters in Table 2.
The dashed lines are power-law fits to the observed data derived by
assuming unit log slope and weighting all points equally.  The observed relations
are
$$ L_{core} =  0.012 \ L_{gal},
                                                            \eqno{(19)}
$$
and
$$ r_b =  0.03 \ r_{e}.
                                                            \eqno{(20)}
$$
The gray areas represent the ranges covered by 
the theoretical predictions in Eqs. (17)
and (18); in plotting Fig. 9a, it is assumed that $M_{core}/M_{gal} =
L_{core}/L_{gal}$, which should be true provided $(M/L)_V$ for the stars
does not vary strongly between 10\asec \ and $r_{e}$.

Observed core radii are within the range predicted by the
models for $M_{\bullet}/M_{gal} = 0.002$, while core luminosities are near
the upper boundary of the predicted range.  The predicted trends as a function
of luminosity are generally matched, although observed core masses may increase as a
steeper-than-unity power of total mass (cf. Eq. 8). 

The Quinlan-Hernquist models confirm the suggestion (Ebisuzaki \et 1991, Makino \&
Ebisuzaki 1996) that mergers of galaxies containing massive BHs 
can generate cores with roughly the size
and luminosity indicated by the observations.  The 
presence of BHs in cores also helps to defend cores against accretion: small
dense satellites accreted by a primary core galaxy may be disrupted 
by the BH before they sink to the center.

Despite these encouraging results, models of core formation by massive BHs
remain uncertain. In particular, (1) the existing simulations do not yet explore 
the full range of BH mass ratios and initial conditions appropriate for merging
galaxies; and (2) initial BH formation has simply been posited in
{\it ad hoc} fashion in all
hot galaxy progenitors (see Haehnelt \& Rees 1993).

There are also further problems to be considered:

\item{(i)} Why do core profiles exhibit weak cusps? Perhaps the 
slow shrinkage of the BH binary naturally forms a cusp, either
because the stars are
flung into elongated orbits by the binary, or in the same way that cusps are
formed when a single central BH grows adiabatically (Peebles 1972; Young 1980;
Quinlan, Hernquist \& Sigurdsson 1995).  However, the 
N-body models of Quinlan \& Hernquist do not show such cusps at present
resolution.  Alternatively, gas infall into
a core previously formed by a BH binary might steepen the profile to create a
cusp (Begelman \et 1980; Young 1980).  Gas is apparently collecting now at the
centers of at least some core galaxies (e.g., M87, Ford \et 1994).

\item{(ii)} What is the relation of nuclei to BHs? Do they signal BHs,
compete with BHs, or possibly feed BHs?  The nucleus in NGC~3115 (Kormendy \et
1996b) has a stellar mass of $\sim 3 \times 10^7$ M$_{\solar}$ crammed into a
tiny volume of radius $\sim$2 pc around a BH that is 50 times more massive.
The stellar density approaches $10^6 {\rm M}_{\solar} {\rm pc}^{-3}$, and
typical orbital velocities exceed 1000 km s$^{-1}$.  It is a puzzle
how stars could have formed in such an
environment, where gas clouds are likely to be colliding at
high velocities while bathed by intense radiation from the BH.
Perhaps the nucleus
and BH formed in different progenitors that later merged. 

\item{(iii)} Our discussion so far has stressed correlations of core and
global properties. In fact there are outliers such as Fornax
A, which has a very small core for its luminosity (Fig. 4a).  Fornax A
is peculiar and is 
probably still in the throes of a major merger (Schweizer 1980).  Perhaps
the inner regions have not yet settled down to their final state, giving us a
clue to the time scales involved in core scouring; 
or gas (as signaled by the copious dust)
may be (re)forming a stellar cusp, although there is no sign of young stars in the
color map (Shaya \et 1996). 

\item{(iv)} BHs appear to be associated with the hot component of spiral
galaxies; late-type spirals such as M33 have little or no central BH (Kormendy
\& McClure 1993). When two late-type galaxies merge, they are believed to form
an elliptical, but this will not have the central BH that is required for core
formation in subsequent hierarchical merging. 

\item{(v)} Most important, if cores are formed by merging binary BHs, why do power-law
galaxies at intermediate magnitudes ($-20.5 > M_V > -22$) not have cores
with size as given by Eq. (18)?  BHs appear to
be just as common in power-law galaxies (Kormendy \& Richstone 1995).
Perhaps power laws can be regenerated
by star formation from fresh gas supplied by the latest merger.  However, to
avoid being ejected by the BH binary, the new stars must form {\it after} the
BH binary shrinks, which poses a timing problem if BHs sink to the center more
slowly than gas. 

\bigskip\bigskip
\centerline{9. SUMMARY AND CONCLUSIONS}
\bigskip

We have assembled inner surface-brightness profiles for 61 dynamically hot
galaxies available in the HST archive as of June 1993.  Fits of the nuker law
(Eq. 2) to deconvolved profiles are used to compute values of break radius
$r_b$ and break surface brightness $I_b$.  These are supplemented with
ground-based data from the literature on rotation, isophote shape, and
velocity dispersion.  These data are used to produce updated versions of the
central parameter diagrams for hot galaxies.
 
The inner surface-brightness profiles of hot galaxies can be divided into two
types as discussed in Paper I. {\it Core} galaxies have a sharp knee or bend
in the profile, akin to the analytic cores of King models or the isothermal
sphere but with a shallow cusp at small radii. {\it Power-law} galaxies have
profiles that are steep and rather featureless in log-log coordinates with no
detectable core at $0\s.1$ resolution.
 
Cores appear only in galaxies brighter than $M_V = -20.5$ (Fig. 4a); 
core size and luminosity are roughly proportional to galaxy size and
luminosity (Eqs. 18 and 19).  Power-law galaxies are fainter than $M_V = -22$.
In the overlap region from $M_V = -20.5$ to $-22$, the two types coexist and
profile morphologies vary widely --- upper limits to core size in some
power-law galaxies are at least 100 times smaller than the core sizes of other
galaxies at the same luminosity.

 The scatter in central properties in the overlap region correlates with
global structure: core galaxies tend to be boxy and rotate slowly, while
power-law galaxies are disky and rotate rapidly. Preliminary evidence suggests
a further correlation with environment in that core
galaxies tend to be found in dense groups and clusters
while bright power laws are preferentially found in the field. 
 
Cores populate a fundamental plane (FP) that is analogous to and roughly
parallel to the global FP for elliptical galaxies.  The scatter about this
plane (in $\log r_b$) is 0.12 dex, about 30\% larger than the analogous
scatter about the global FP.  Some of this extra scatter may come from massive
BHs, which may inflate central velocity dispersions in some galaxies.
 
A set of 
self-consistent scaling relations for core galaxies is presented that expresses core
size, density, and other quantities as a function of $L_V$
(Eqs. 3--10).  These scaling laws are projections of the FP.
A major conclusion 
is that small hot galaxies are much denser than large ones, by a factor of up
to 1000 at a radius of 10 pc.
 
The last part of the paper attempts to relate the central parameter
relations of hot galaxies to the process of galaxy formation and evolution.
We suggest that the presence of dense power-law centers, disky isophotes, and
rapid rotation in low-luminosity galaxies all point to their formation via
{\it dissipative, gas-rich mergers}.  The analogous arguments about core galaxies
are less clear: the boxy isophote shape and slow rotation of these luminous
objects suggest formation by {\it dissipationless}
mergers, but cores may be difficult to form and maintain in such events.
For example, core galaxies seem at present to be accreting small dense
satellites in sufficient numbers to fill in their low-density cores, at
least if the satellites survive their orbital decay 
to the center, an issue that is still
in dispute. 
 
We explore an alternative model for core formation based on merging
BHs.  The model assumes that BH binaries are formed in galaxy mergers; the
binary orbit decays by dynamical friction, ejecting stars from the center of
the merger remnant, enlarging any previous core, and scouring out a new one
where none existed. Simulations of this process by Quinlan \& Hernquist (1997) yield a
reasonable match to the radii and masses of observed cores if every hot galaxy
contains a central BH of average mass $M_{\bullet} = 0.002~ M_{gal}$.  This
value for $M_{\bullet}$ is consistent with BH mass estimates in AGNs and
local BHs (see Appendix B).  Whether or not BHs
are the dominant agent in creating cores, their role in shaping the central
structure of hot galaxies is likely to be significant if they are as common
and as massive as recent estimates suggest.
 
The main goal of this paper is to explore systematic trends in the central
structure of hot galaxies and the possible relations between their present
central structure and their formation history.  By strengthening the link
between the central structures of hot galaxies and their global properties
such as luminosity, shape, and rotation, {\it HST} has helped to open an
important new window on galaxy formation.

\bigskip
We would especially like to thank Martin Weinberg and Gerry Quinlan for
informative discussions on satellite capture and survival and for providing
pre-publication accounts of their calculations.  John Tonry provided distance
estimates to many galaxies in advance of publication, for which we are very
grateful.  The referee made several very helpful suggestions.
During the last five years, our team has been hosted by Prof. James
Westphal of Caltech, the Institute for Astronomy at the University of Hawaii,
the Observatories of the Carnegie Institution of Washington, the Institute for
Theoretical Physics at UCSB, the National Optical Astronomy Observatories, the
Aspen Center for Physics, and the Fields Institute for Research in
Mathematical Sciences at the University of Toronto.  We thank them for their
gracious hospitality.  Our collaboration was supported by {\it HST} data
analysis funds through GO grants GO-2600.01.87A and GO-06099.01-94A, by NASA
grant NAS-5-1661 to the WFPC1 IDT, and by grants from NSERC.

\vfill\eject 

\centerline{APPENDIX A: THE NATURE OF POWER LAWS AND THE}
\centerline{FREQUENCY OF INNER DISKS IN POWER-LAW GALAXIES}
\bigskip

Jaffe \et (1994) and Ferrarese \et (1994), like us, divide
hot galaxies into two types based on inner surface-brightness
profile.  Those called by us {\it cores} 
with a strong break and low central surface
brightness they
term Type I, and those called by us {\it power-laws}
with no break and high central surface they call Type II.  There is
no discrepancy between us as to division into classes 
based on profile shape.

Jaffe {\it et al.} go on to identify core galaxies 
in a general way
with slowly rotating boxy galaxies, and power-law
galaxies with rotating disky galaxies.  This
distinction resembles ours but differs in important details.
For example, 
Jaffe {\it et al.}  envision that the centers of hot galaxies 
either have or do not have small {\it inner disks}.  Such disks
{\it seen edge-on} are what create the high surface brightness power-law
profiles of Type II galaxies.  These inner disks are furthermore
associated with the global, outer disks of rotating disky
galaxies.  It is the frequent association between inner disks and
outer disks that creates the link between power-law profiles
and disky rotating galaxies in their picture.

Jaffe {\it et al.} believe that the high surface brightness
profiles of power-law galaxies are produced {\it only} when an inner
disk is seen edge-on.  
Specifically they state: ``Most of the characteristics$\ldots$that discriminate
Type I from Type II are explained by disk components seen at high inclination
angles.  For example, the higher central surface brightness in Type II systems
is caused by the nuclear disk seen close to edge-on$\ldots$If one of these
[disky] galaxies [i.e., a Type II] were viewed face on, it would appear much more like a
Type I galaxy."  Thus, their view is that observed profile type is due to a
combination of intrinsic properties plus viewing aspect.  All power-laws
have inner disks -- they are the disks that happen to be 
seen edge-on.  Core-type profiles on the other hand
are a mixture; many are intrinsic cores that lack inner disks, while some 
fraction are disks seen face-on that are masquerading as cores.

This interpretation of cores versus power laws differs importantly from our
own.  Our view is that the difference
between core and power-law profiles is intrinsic to the {\it hot} stellar component 
and has no direct connection with a disk, whether seen edge-on or face-on.
Power laws remain power laws at any viewing angle,
as do cores.  Paper I presented
initial arguments against the edge-on disk explanation for power laws.
We have 
since undertaken a more comprehensive comparison with power-law galaxies in
the present data set.  Briefly, we find that the small sample of Virgo
galaxies analyzed by Jaffe {\it et al.} was abnormally 
dominated by late-type edge-on S0 galaxies.  Nearly every power-law
galaxy they detected was such an object.  A high frequency of edge-on
{\it inner} disks in such a sample is therefore understandable.  Our sample is larger and
contains many power-law ellipticals that are not flattened and show
no sign of either an inner or an outer disk.  This and other evidence to
be described leads us to conclude that power-laws are independent of
inner disks and are thus a feature of the hot stellar component alone.

We carefully examined deconvolved V-band images of all 61 galaxies in the present
sample.  Thirteen of the 14 Jaffe {\it et al.}  galaxies were available in the
archive, and we looked at all of them.  Seven of these were admitted
into our sample (they are included in Table 1).  The remaining six galaxies
were rejected for the following reasons: (1) too much dust to
derive a reliable surface-brightness profile or class the object as a core or
power law (NGC~4261, NGC~4342, NGC~4374, NGC~4476); (2) a potential double
nucleus and unclassifiable profile (NGC~4473); (3) no clear spheroidal
component (NGC~4550); and (4) interfering spiral arms (NGC~4476).  Compared to
our sample, the Jaffe {\it et al.} power-law galaxies are much later in
type, diskier, more edge-on, and more subject to dust and other peculiarities
that potentially interfere with reliable measurement of the spheroid profile.
 
If the edge-on disk interpretation were correct, then all or most 
power-law galaxies in our sample should show evidence of disks.  To
test this, we inspected each image within a 10\asec \ radius for an inner disk
comprised of either stars or dust.  We also looked for spiral arms, which we
took as another indicator of a disk.  We visually assessed isophote shape as a
function of radius, checking for changes and comparing to measured values of
$a_4$.  We tried to correlate changes in ellipticity and shape
with kinks or ``ledges" in the brightness profile --- such
correlations might signify the edge of a disk.  We also subtracted the profile fits
given in Paper I and looked for signs of a residual disk.  Table 4 summarizes
the results of this visual inspection.  On the basis of this evidence, we
assigned a final score to each galaxy indicating the likelihood of an {\it inner}
disk.  The values are 0 (no sign of a disk), 1 (possible disk), 2 (probable
disk), and 3 (disk plainly visible).  These scores are given in the table,
along with comments.

The results of these efforts are summarized in Fig. 10, which plots inner
disk score versus ellipticity for power-law galaxies.  The hypothesis that
{\it all} power-law objects have edge-on inner disks seems unlikely because:
(1) Half the sample shows little or no evidence of any inner
disk, including several highly flattened objects. (2) Power-law galaxies are
not concentrated at high ellipticities, in contrast to the sample of Jaffe
{\it et al}. (3) Most objects with inner disks are bulges that also have
outer disks (types
S0, Sa, or Sb), a point also made by Jaffe {\it et al}.  However, this weakens
the case for ubiquitous 
inner disks in power-law {\it ellipticals} because they lack such outer
disks. (4) One power-law galaxy, NGC~3599, shows face-on spiral structure,
showing conclusively that it cannot be edge-on. (5) If known
bulges are excluded, the ellipticity distribution of the remaining power-law
ellipticals is nearly the same as that of core ellipticals (which lie within
the rectangle in Fig. 10 but are not plotted individually), and neither type
shows much evidence for inner disks.  Thus the evidence for inner disks is
weak in both power-law and core ellipticals.

A less restrictive hypothesis, not put forward by Jaffe {\it et al.},
 is that power laws are associated with a high
surface-brightness inner disk, period, whether seen {\it either} edge-on or
face-on.  This also seems unlikely because there are
several flattened power-law ellipticals that must be close to edge-on yet show
no sign of a disk (point 1 above).  
The kinematic properties of M~31 and M~32 are also relevant here.
Both of these would be typical power laws if seen at a distance, yet both are hot and
slowly rotating at radii of a few arcsec, the claimed size of inner disks in
other power-law galaxies.  In neither galaxy is there any hint from 
kinematics that the high central surface brightness is associated with a disk.

To summarize, present evidence does not favor the ubiquitous presence of high
surface-brightness {\it inner} disks in power-law galaxies, whether edge-on or not,
though such disks are certainly present in some cases.  Rather we believe that the
difference between core and power-law profiles more probably reflects an
intrinsic difference in the {\it spheroidal} light distribution between the two
types.  Finally we note cautiously that one of our low-disk-score galaxies,
NGC~3377 (disk score = 1), has since revealed a dust disk in recent WFPC2
images.  Final conclusions about the frequency of inner disks in ellipticals
and bulges should therefore await a new body of high-quality WFPC2 images.

\bigskip\bigskip
\centerline{APPENDIX B: THE MEAN BH MASS PER HOT GALAXY}
\bigskip

The following argument adapted from Tremaine (1997) summarizes the evidence
for the frequency and masses of BHs in the centers of hot galaxies.

The integrated comoving energy density in quasar light (as emitted) is
(Chokshi \& Turner 1992)
$$ u = 1.3 \times 10^{-15} {\rm \ erg \ cm}^{-3}, \eqno{(21)} $$ 
independent of $H_0$ and $\Omega$.  If this energy is produced by burning fuel
with an assumed efficiency $\epsilon \equiv \Delta E/(\Delta Mc^2)$, then the
mean mass density of dead quasars must be at least (So\l tan 1982, Chokshi \&
Turner 1992)
$$ \rho_{\bullet} = {u \over {\epsilon c^2}} = 
   {2.2 \times 10^5} {\biggl({0.1 \over \epsilon} \biggr)} {\rm M}_{\solar}
   {\rm Mpc}^{-3},   \eqno{(22)}  $$
assuming that the Universe is homogeneous and transparent.

The mass of a dead quasar may be written
$$ M_{\bullet} = {{L_Q\tau} \over {\epsilon c^2}} =
     {7 \times 10^8{\rm M}_{\solar}} {\biggl({{L_Q} \over {10^{12} {\rm L}_{\solar}}} \biggr)}
     {\biggl({\tau \over {10^9 {\rm y} }}\biggr)}
     {\biggl({0.1 \over \epsilon} \biggr)},   \eqno{(23)}  $$
where $L_Q$ is the quasar luminosity and $\tau$ is its lifetime.  An upper
limit to the lifetime is the evolution timescale for the quasar population
as a whole, $\sim 10^9$ y; however, upper limits to BH masses
in nearby galaxies and direct estimates of the BH masses in
AGNs both suggest that the actual masses and lifetimes are smaller by a factor
10-100 (Haehnelt \& Rees 1993), which implies $M_{\bullet} = 10^7$--$10^8 
{\rm M}_{\solar}$.

To focus discussion, we adopt a strawman model in which
a fraction $f$ of all galaxies contains a central BH, and BH mass
is proportional to galaxy luminosity.  Thus $M_{\bullet} = 
\Upsilon L$, where $\Upsilon$ is the (black hole) to (galaxy)
mass-to-light ratio.  The luminosity density of galaxies is 
$j = 1.5 \times 10^8{\rm L}_{\solar}{\rm Mpc}^{-3}$ in the blue
band (Efstathiou \et 1988, adjusted to our Hubble constant
of $H_0$ = 80 \kms Mpc$^{-1}$).  Thus
$$ \Upsilon = {\rho_{\bullet} \over {fj}} = 
   {0.0015 \over f}{\biggl({0.1 \over \epsilon} \biggr)} 
   {{\rm M}_{\solar} \over {\rm L}_{\solar}}.   \eqno{(24)}  $$
This value is an average over the light of all local galaxies.
However, if we assume that massive BHs are found chiefly in
the centers of hot galaxies (Kormendy
\& Richstone 1995), the above number can be converted
to the BH mass-to-light ratio {\it per hot component} by noting that approximately
30\% of the local B-band light is emitted by such components (Schechter
\& Dressler 1987).  Correcting for this and converting
to the V band yields
$$ \Upsilon^{h}_V = {0.004 \over f_h}{ \biggl({0.1 \over \epsilon} \biggr)} 
   {{\rm M}_{\solar} \over {\rm L}_{\solar}},   \eqno{(25)}  $$
where $\Upsilon^{h}_V$ is now the estimated ratio $M_{\bullet}/L_V$ 
per hot component and $f_h$ is the fraction of {\it hot galaxies}
with BHs.

A second estimate of $\Upsilon^{h}_V$ from quasars comes from dividing
the typical dead quasar mass derived above, $M_{\bullet} \approx 10^{7.5}{\rm
M}_{\solar}$, by the typical luminosity of a bright hot component, $8.5 \times
10^9{\rm L}_{\solar}$ (Binggeli, Sandage \& Tammann 1988, adjusted to $H_0$ =
80 \kms Mpc$^{-1}$).  The result is $\Upsilon^{h}_V \approx 0.004$.
Consistency with Eq. (25) then requires $f_h \approx 1$ if $\epsilon
\approx 0.1$, or that most or all hot galaxies contain a BH.

A final method for estimating $\Upsilon^{h}_V$ uses individual BH masses for
6 moderately well established BHs in nearby hot galaxies\footnote{$^{13}$} {The
Galaxy and NGC 4258 are omitted for lack of accurate luminosities of their
hot components,
and NGC 3115 uses the new value of $M_{\bullet} = 2 \times 10^9 {\rm
M}_{\solar}$ from Kormendy \et (1996b).}  (Kormendy \& Richstone 1995, Table
1). Using an estimate of global stellar $M/L_V$ based on nuker-law fits as
described in Table 3\footnote{$^{14}$} {For the present purpose, the mass fits
were renormalized to fit $\sigma$ at 10\asec \ using the ratio $R_{\sigma} =
\sigma(10\asec)/\sigma_0$ from Table 2.  This was done to avoid possible
contamination of $\sigma_0$ by a BH.  The logarithmically averaged $(M/L_V)$
for the 6 galaxies is 4.0 in solar units ($H_0 = 80$ \kms Mpc$^{-1}$).}
yields the logarithmic mean value $\Upsilon^{h}_V = 0.016$.  This is in
reasonable agreement with $\Upsilon^{h}_V = 0.004$ from quasars in
view of the likelihood that these best BH candidates are more massive than
average.

For further discussion, we assume that every hot galaxy contains a BH and
adopt for $\Upsilon^{h}_V$ the logarithmic mean of the quasar 
and BH values:
$$               \Upsilon^{h}_V = 0.008.    \eqno{(26)}$$
The corresponding value of $M_{\bullet}/M_{gal}$ is then
$$               M_{\bullet}/M_{gal} = 0.002,    \eqno{(27)}$$
where $M_{gal}$ comes from $L_V (M/L)_V$ and $(M/L)_V$ is the 
above-mentioned mean global mass-to-light ratio for the
6 candidate BH galaxies.  This is the value of $M_{\bullet}/M_{gal}$ 
per hot component adopted in Section 8.

\vfill\eject

\centerline{REFERENCES}

\bigskip\noindent
\ref{Balcells, M., \& Quinn, P. J. 1990, ApJ, 361, 381}

\medskip\noindent
\ref{Bardeen, J. M., Bond, J. R., Kaiser, N., \& Szalay, A. S.
1986, ApJ, 304, 15}

\medskip\noindent
\ref{Barnes, J. E. 1988, ApJ, 331, 699}

\medskip\noindent
\ref{Barnes, J. E. 1992, ApJ,  393, 484}

\medskip\noindent
\ref{Barnes, J. E. 1996, in New Light on Galaxy Evolution,
IAU Symposium No. 171, eds. R. Bender \&
R. L. Davies (Kluwer, Dordrecht), 191}

\medskip\noindent
\ref{Barnes, J. E., \&  Hernquist, L. 1991, ApJ,  370, L65}

\medskip\noindent
\ref{Barnes, J. E., \& Hernquist, L. 1992, ARA\&A, 30, 705}

\medskip\noindent
\ref{Barnes, J. E., \& Hernquist, L. 1996, ApJ, 471, 115}

\medskip\noindent
\ref{Begelman, M. C., Blandford, R. D., \& Rees, M. J. 1980, Nature, 287, 307}

\medskip\noindent
\ref{Bender, R. 1988, A\&A,  193, L7  }


\medskip\noindent
\ref{Bender, R., Burstein, D., \& Faber, S.M. 1992, ApJ,  399, 462}


\medskip\noindent
\ref{Bender, R, \& M\"ollenhoff, C. 1987, A\&A, 177, 71}

\medskip\noindent
\ref{Bender, R., \& Nieto, J.-L. 1990, A\&A,  239, 97 }

\medskip\noindent 
\ref{Bender, R., Saglia, R. P., \& Gerhard, O. E. 1994, MNRAS, 269, 785 } 

\medskip\noindent
\ref{Bender, R., \& Surma, P. 1988, A\&A, 258, 250}

\medskip\noindent
\ref{Bender, R., \& Surma, P. 1995, A\&A, 298, 405}

\medskip\noindent
\ref{Bender, R., Surma, P., D\"obereiner, S., M\"ollenhoff, C.,
\& Madejsky, R. 1989, A\&A, 217, 35}

\medskip\noindent
\ref{Bertola, F., Capaccioli, M., Galletta, G., 
\& Rampazzo, R. 1988, A\&A,  192, 24 }

\medskip\noindent
\ref{Binggeli, B., \& Cameron, L. M. 1993,  A\&AS,  98, 297 } 

\medskip\noindent
\ref{Binggeli, B., Sandage, A. R., \& Tammann, G. A. 1985, AJ,  90, 168 } 

\medskip\noindent
\ref{Binggeli, B., Sandage, A. R., \& Tammann, G. A. 1988, ARA\&A, 26, 509}

\medskip\noindent
\ref{Binney, J., Davies, R. L., \& Illingworth, G. D. 1990,   ApJ,  361, 78}

\medskip\noindent
\ref{Binney, J., \& Petrou, M.  1985,   MNRAS,  214, 449 }

\medskip\noindent
\ref{Binney, J. \& Tremaine, S. 1987, Galactic Dynamics (Princeton 
University Press,  Princeton), 514 }

\medskip\noindent
\ref{Blumenthal, G., Faber, S. M., 
Primack, J., \& Rees, M. J. 1984,  Nature, 311, 517}

\medskip\noindent
\ref{Boroson, T. 1981, ApJS,  46, 177 }

\medskip\noindent
\ref{Bosma, A. Smith, R. M., \& Wellington, K. J. 1985, MNRAS, 212, 301} 

\medskip\noindent
\ref{Bower, G., \et 1997, in preparation} 

\medskip\noindent
\ref{Burkhead, M. 1986, AJ,  128, 465 } 

\medskip\noindent
\ref{Burstein, D. 1979,  ApJ,  234, 435 }

\medskip\noindent
\ref{Bushouse, H. A. 1987, ApJ, 320, 49}

\medskip\noindent
\ref{Byun, Y.-I., \et 1996, AJ, 111, 1889 (Paper II)}

\medskip\noindent
\ref{Capaccioli, M., Held, E. V., \& Nieto, J.-L. 1987, AJ, 94, 1519 } 

\medskip\noindent
\ref{Carlberg, R. 1986, ApJ,  310, 593}

\medskip\noindent
\ref{Chokshi, A., \& Turner, E. L. 1992, MNRAS, 259, 421}

\medskip\noindent
\ref{Condon, J. J., Huang, Z.-P., Yin, Q. F., \& Thuan, T. X.
1991, ApJ, 378, 65 }

\medskip\noindent
\ref{Crane, P. \et 1993, AJ, 106, 1371}


\medskip\noindent
\ref{Davies, R. L., \& Birkinshaw, M. 1988, ApJS, 68, 
409 } 

\medskip\noindent
\ref{Davies, R. L., Efstathiou, G., Fall, S. M, Illingworth, G. D., 
\& Schechter, P. L. 1983, ApJ,  266, 41 }

\medskip\noindent
\ref{Davies, R. L., \& Illingworth, G. D. 1983, ApJ,  266, 516 } 

\medskip\noindent
\ref{de Vaucouleurs, G. 1958,  ApJ,  128, 465 } 

\medskip\noindent
\ref{de Vaucouleurs, G., de Vaucouleurs, A., Corwin, H. G., Buta,
R. J., Paturel, G., \& Fouqu\'e, P. 1991, Third Reference
Catalogue of Bright Galaxies (Springer, New York) (RC3) }

\medskip\noindent
\ref{de Vaucouleurs, G., de Vaucouleurs, A., \& Corwin, H. G., 
1976, Second Reference
Catalogue of Bright Galaxies (Univ. of Texas Press, Austin) (RC2) } 

\medskip\noindent
\ref{de Zeeuw, T. \& Franx, M. 1991, ARA\&A, 29, 239}

\medskip\noindent
\ref{Djorgovski, S., \& Davis, M. 1987, ApJ,  313, 59}

\medskip\noindent
\ref{Dressler, A., Lynden-Bell, D., Burstein, D., Davies, R. L., Faber, S. M.,
Terlevich, R., \& Wegner, G. 1987, ApJ,  313, 42}

\medskip\noindent
\ref{Dressler, A. \& Richstone, D. O. 1988,  ApJ,  324, 701 }

\medskip\noindent
\ref{Dressler, A. \& Richstone, D. O. 1990,  ApJ,  348, 120 } 


\medskip\noindent
\ref{Dubinski, J. 1997, in preparation } 

\medskip\noindent
\ref{Ebisuzaki, T., Makino, J., \& Okumura, S. K. 1991,  Nature, 354, 212}

\medskip\noindent
\ref{Efstathiou, G., Ellis, R. S., \& Carter, D. 1980, MNRAS, 193, 931} 

\medskip\noindent
\ref{Efstathiou, G., Ellis, R. S., \& Carter, D. 1982, MNRAS, 201, 975  } 

\medskip\noindent
\ref{Efstathiou, G., Ellis, R. S., \& Peterson, B. A. 1988, MNRAS, 232, 431}

\medskip\noindent
\ref{Faber, S. M. 1973, ApJ, 179, 423}

\medskip\noindent
\ref{Faber, S. M., Dressler, A., Davies, R. L., Burstein, D., Lynden-Bell, D.,
Terlevich, R., \& Wegner, G. 1987, in Nearly Normal Galaxies: From the
Planck Time to the Present, ed. S. M. Faber (Springer, New York), 175}

\medskip\noindent
\ref{Faber, S. M., Wegner, G., Burstein, D., Davies, R. L., Dressler, A.,
Lynden-Bell, D., \& Terlevich, R. J. 1989,  ApJS, 69, 763 } 

\medskip\noindent
\ref{Farouki, R. T., Shapiro, S. L., \& Duncan, M. J. 1983, ApJ, 265, 597}

\medskip\noindent
\ref{Ferrarese, L., van den Bosch, F. C., Ford, H. C., Jaffe, W., \&
O'Connell, R. W. 1994, AJ, 108, 1598}

\medskip\noindent
\ref{Ferrarese, L., Ford, H. C., Jaffe, W., 1996, ApJ, 470, 444}

\medskip\noindent
\ref{Fisher, D., Illingworth, G. D., \& Franx, M. 1995,
ApJ,  438, 539 } 

\medskip\noindent
\ref{Forbes, D. A., Franx, M., \& Illingworth, G. D. 1994,  ApJ,
428, L49 }

\medskip\noindent
\ref{Forbes, D. A., Franx, M., \& Illingworth, G. D. 1995, AJ, 109, 1988}

\medskip\noindent
\ref{Ford, H. C., \et 1994, ApJ, 435, L27}

\medskip\noindent
\ref{Franx, M., \& Illingworth, G. D. 1988, ApJ,  327, L55}

\medskip\noindent
\ref{Franx, M., Illingworth, G. D., \& Heckman, T. 1989a,   ApJ,  344, 613}

\medskip\noindent
\ref{Franx, M., Illingworth, G. D., \& Heckman, T. 1989b, AJ,  98, 538}
 
\medskip\noindent
\ref{Fried, J. W., \& Illingworth, G. D. 1994, AJ,  107, 992 }

\medskip\noindent
\ref{Gebhardt, K. \et 1996, AJ, 112, 105 (Paper III)}

\medskip\noindent
\ref{Gebhardt, K. \et 1997, in preparation }

\medskip\noindent
\ref{Gonzalez, J. J. 1993,  Ph.D. thesis, University of California, Santa Cruz}
 
\medskip\noindent
\ref{Goudfrooij, P., Hansen, L., J\o rgensen, H. E., N\o rgaard-Nielsen, 
H. U., de Jong, T., \& van den Hoek, L. B. 1994, A\&AS, 104, 179}

\medskip\noindent
\ref{Grillmair, C. J., \et 1994, AJ, 108, 102}

\medskip\noindent
\ref{Haehnelt, M. G., \& Rees, M. J. 1993, MNRAS, 263, 168}

\medskip\noindent
\ref{Harms, R. J., \et 1994, ApJ, 435, L35}


\medskip\noindent
\ref{Hernquist, L. 1993, ApJ,  409, 548}

\medskip\noindent
\ref{Hernquist, L., \& Barnes, J. E. 1991, Nature, 354, 210}

\medskip\noindent
\ref{Hoessel, J G., Gunn, J. E., \& Thuan, T. X. 1980, ApJ, 241, 486}

\medskip\noindent
\ref{Jaffe, W., Ford, H. C., O'Connell, R. W., van den Bosch, F. C., \&
Ferrarese, L. 1994, AJ, 108, 1567}

\medskip\noindent
\ref{Jaffe, W., Ford, H. C., Ferrarese, L., van den Bosch, F. C., \&
O'Connell, R. W. 1996, ApJ, 460, 214}

\medskip\noindent
\ref{Jarvis, B. J., \& Freeman, K. C. 1985,  ApJ,  295, 324 }

\medskip\noindent
\ref{Jedrzejewski, R. I., \& Schechter, P. L. 1988,  ApJ, 330, L87 }

\medskip\noindent
\ref{Jedrzejewski \& Schechter, P. L. 1989,  AJ, 98, 147 }

\medskip\noindent
\ref{Kauffmann, G. 1996, MNRAS, 281, 487 }

\medskip\noindent
\ref{Kennicutt, R. C., Keel, W. C., van der Hulst, J. M.,
Hummel, E., \& Roettiger, K. A. 1987, AJ, 93, 1011}

\medskip\noindent
\ref{Kent, S. M., 1983, ApJ, 266, 562 }

\medskip\noindent
\ref{Kormendy, J. 1977,  ApJ,  217, 406  }

\medskip\noindent
\ref{Kormendy, J. 1982a, in Morphology and Dynamics of Galaxies, eds.
L. Martinet \& M. Mayor (Sauverny: Geneva Observatory), 113}

\medskip\noindent
\ref{Kormendy, J. 1982b, ApJ,  257, 75 }

\medskip\noindent
\ref{Kormendy, J. 1984, ApJ,  287, 577}

\medskip\noindent
\ref{Kormendy, J. 1985,  ApJ,  295, 73}

\medskip\noindent
\ref{Kormendy, J. 1987a, in IAU Symposium 127, Structure and Dynamics 
of Elliptical Galaxies, ed. T. de Zeeuw (Reidel, Dordrecht), 17}

\medskip\noindent
\ref{Kormendy, J. 1987b, in Nearly Normal Galaxies: From the
Planck Time to the Present, ed. S. M. Faber (Springer, New York), 163}

\medskip\noindent
\ref{Kormendy, J. 1988, ApJ,  325, 128 }

\medskip\noindent
\ref{Kormendy, J. 1989, ApJ,  342, L63}

\medskip\noindent
\ref{Kormendy, J., \& Bender, R. 1996, ApJ, 464, L119}

\medskip\noindent
\ref{Kormendy, J., \& Djorgovski, S. 1989, ARA\&A, 27, 235}

\medskip\noindent
\ref{Kormendy, J., Dressler, A., Byun, Y.-I., Faber, S. M., Grillmair, C.,
Lauer, T. R., Richstone, D. O., \& Tremaine, S. 1994, in ESO/OHP Workshop on
Dwarf Galaxies, eds. G. Meylan \& P. Prugniel (Garching: ESO), 147}

\medskip\noindent
\ref{Kormendy, J., \& Illingworth, G. 1983, ApJ,  265, 632}

\medskip\noindent
\ref{Kormendy, J., \& McClure, R. D. 1993, AJ,  105, 1793}

\medskip\noindent
\ref{Kormendy, J., \& Richstone, D. O. 1992, ApJ,  393, 559}

\medskip\noindent
\ref{Kormendy, J, \& Richstone, D. 1995,  ARA\&A, 33, 581}

\medskip\noindent
\ref{Kormendy, J., \& Sanders, D. B. 1992, ApJ, 390, L53}

\medskip\noindent
\ref{Kormendy, J., \& Westpfahl, D. J. 1989,   ApJ,  338, 752 }

\medskip\noindent
\ref{Kormendy, J., \et 1996a, in New Light on Galaxy Evolution,
IAU Symposium No. 171, eds. R. Bender \& 
R. L. Davies (Kluwer, Dordrecht), 105}

\medskip\noindent
\ref{Kormendy, J., \et 1996b, ApJ, 459, L57}

\medskip\noindent
\ref{Kormendy, J., \et 1996c, ApJ, 473, L91}

\medskip\noindent
\ref{Kormendy, J., \et 1997, ApJ, 482, L139}

\medskip\noindent
\ref{Larson, R. B. 1974a, MNRAS, 166, 585}

\medskip\noindent
\ref{Larson, R. B. 1974b, MNRAS, 169, 229}

\medskip\noindent
\ref{Lauer, T. R. 1983,  Ph.D. Thesis, University of California, Santa Cruz}

\medskip\noindent
\ref{Lauer, T. R. 1985a, ApJ, 292, 104}

\medskip\noindent
\ref{Lauer, T. R. 1985b, MNRAS, 216, 429}

\medskip\noindent
\ref{Lauer, T. R. 1988, ApJ,  325, 49}

\medskip\noindent
\ref{Lauer, T. R., \et 1991, ApJ,  369, L41   }

\medskip\noindent
\ref{Lauer, T. R., \et 1992a, AJ,  104, 552   }

\medskip\noindent
\ref{Lauer, T. R., \et 1992b, AJ,  103, 703   }

\medskip\noindent
\ref{Lauer, T. R., \et 1993, AJ, 106, 1436 }

\medskip\noindent
\ref{Lauer, T. R., \et 1995, AJ, 110, 2622 (Paper I)}

\medskip\noindent
\ref{Lauer, T. R., \et 1996, ApJ, 471, L79}

\medskip\noindent
\ref{Lauer, T. R., \et 1997, in preparation }

\medskip\noindent
\ref{Leech, K. J., Penston, M. V., Terlevich, R., Lawrence, A.,
Rowan-Robinson, M., \& Crawford, J. 1989, MNRAS, 240, 349}

\medskip\noindent
\ref{Lugger, P. M., Cohn, H., Cederbloom, S. E., Lauer, T. R., 
\et 1992, AJ, 104, 83 }

\medskip\noindent
\ref{Lynden-Bell, D., Faber, S. M., Burstein, D., Davies, R. L., Dressler, A.,
Terlevich, R. J., \& Wegner, G. 1988,   ApJ,  326, 19}

\medskip\noindent
\ref{Makino, J., \& Ebisuzaki, T. 1996, ApJ, 465, 527}

\medskip\noindent
\ref{Merritt, D. 1985, ApJ,  289, 18}

\medskip\noindent
\ref{Mihos, J. C., \& Hernquist, L. 1994, ApJ,  437, L47}


\medskip\noindent
\ref{Negroponte, J., \& White, S. D. M. 1983, MNRAS, 205, 1009}

\medskip\noindent
\ref{Nieto, J.-L. 1988, Bol. Acad. Nac. Cine. Cordoba, 58, 239 }


\medskip\noindent
\ref{Nieto, J.-L., \& Bender, R. 1989, A\&A, 215, 266}

\medskip\noindent
\ref{Nieto, J.-L., Bender, R., \& Surma, P. 1991a, A\&A, 244, L37}

\medskip\noindent
\ref{Nieto, J.-L., Bender, R., Arnaud, J., \& Surma, P. 1991b, A\&A, 244, L25}

\medskip\noindent
\ref{Nieto, J.-L., Poulain, P., Davoust, E., Rosenblatt, P. 1991c,
A\&AS, 88, 559 }

\medskip\noindent
\ref{Ostriker, J. P. 1980, Comm. Astrophys., 8, 177}

\medskip\noindent
\ref{Peebles, P. J. E. 1972, Gen. Rel. Grav. 3, 63 }

\medskip\noindent
\ref{Peletier, R. F. Davies, R. L. Illingworth, G. D., Davis, L. E.  1990,
AJ, 100, 1091}

\medskip\noindent
\ref{Quinlan, G. D. 1997, New Astron. 1, 35}

\medskip\noindent
\ref{Quinlan, G. D., \& Hernquist, L.  1997, submitted to New Astronomy }

\medskip\noindent
\ref{Quinlan, G. D., Hernquist, L., \& Sigurdsson, S. 1995, ApJ, 440, 554} 

\medskip\noindent
\ref{Minske, K., \& Richstone, D. O. 1997, in preparation} 

\medskip\noindent
\ref{Sanders, D. B., Soifer, B. T., Elias, J. H., Madore, B. F.,
Mathews, K., Neugebauer, G., \& Scoville, N. Z. 1988,
ApJ, 325, 74}

\medskip\noindent
\ref{Schechter, P. L., \& Dressler, A. 1987, AJ 94, 563}

\medskip\noindent
\ref{Schweizer, F. 1980, ApJ, 237, 303}

\medskip\noindent
\ref{Schweizer, F. 1986, Science, 231, 227}

\medskip\noindent
\ref{Schweizer, F., Seitzer, P., Faber, S. M., Burstein, D.,
Dalle Ore, C. M., \& Gonzalez, J. J. 1990, ApJ, 364, L33}

\medskip\noindent
\ref{Scorza, C., \& Bender, R. 1995, A\&A, 293, 20 }

\medskip\noindent
\ref{Scoville, N. Z., Sargent, A. I., Sanders, D. B., \&
Soifer, B. T. 1991, ApJ, 366, L5}

\medskip\noindent
\ref{Shaya, E. J., \et 1996, AJ, 111, 2212}

\medskip\noindent
\ref{Simien, F., \& de Vaucouleurs, G. 1986, ApJ,  302, 564 }

\medskip\noindent
\ref{Solomon, P. M., Downes, D., \& Radford, S. J. E. 1992,
ApJ, 387, L55}

\medskip\noindent
\ref{So\l tan, A. 1982, MNRAS, 200, 115}


\medskip\noindent
\ref{Tonry, J. 1984, ApJ,  283, L27  }

\medskip\noindent
\ref{Toomre, A. 1977,  in The Evolution of Galaxies \& Stellar Populations,
eds. B. M. Tinsley \& R. B. Larson (Yale, New Haven), 401}

\medskip\noindent
\ref{Tremaine, S. 1976, ApJ, 203, 345}

\medskip\noindent
\ref{Tremaine, S. 1997, in Unsolved Problems in Astrophysics,
eds. J. N. Bahcall and J. Ostriker (Princeton University Press, Princeton), 
p. 137 }

\medskip\noindent
\ref{van den Bosch, F. C., Ferrarese, L., Jaffe, W., Ford, H. C., \&
O'Connell, R. W. 1994, AJ, 108, 1579 }

\medskip\noindent
\ref{van der Marel, R. P. 1991, MNRAS, 253, 710 }

\medskip\noindent
\ref{van der Marel, R. P., Rix, H.-W., Carter, D., Franx, M., White, S. D. M.,
de Zeeuw, T. 1994, MNRAS, 268, 521 }

\medskip\noindent
\ref{van der Marel, R. P., de Zeeuw, T., Rix, H.-W., \& Quinlan, G. D.
1997, Nature, 385, 610 \et 1996,   }

\medskip\noindent
\ref{Weil, M.  \& Hernquist. L. 1996, ApJ, 460, 101}

\medskip\noindent
\ref{Weinberg, M. D. 1994, AJ, 108, 1398}

\medskip\noindent
\ref{Weinberg, M. D. 1997, ApJ, 478, 435}

\medskip\noindent
\ref{White, S. D. M., \& Rees, M. J. 1978, MNRAS, 183, 341}

\medskip\noindent
\ref{Whitmore, B., McElroy, D. B., \& Tonry, J. L. 1985, ApJS, 59, 1 }

\medskip\noindent
\ref{Young, P. J. 1980, ApJ,  242, 1232}

\medskip\noindent
\ref{Young, P. J., Westphal, L. A., Kristian, J., Wilson, C. P., \& Landauer,
F. P.  1978,  ApJ,  221, 721 }

\vfill\eject

\centerline{FIGURES}

\bigskip\noindent
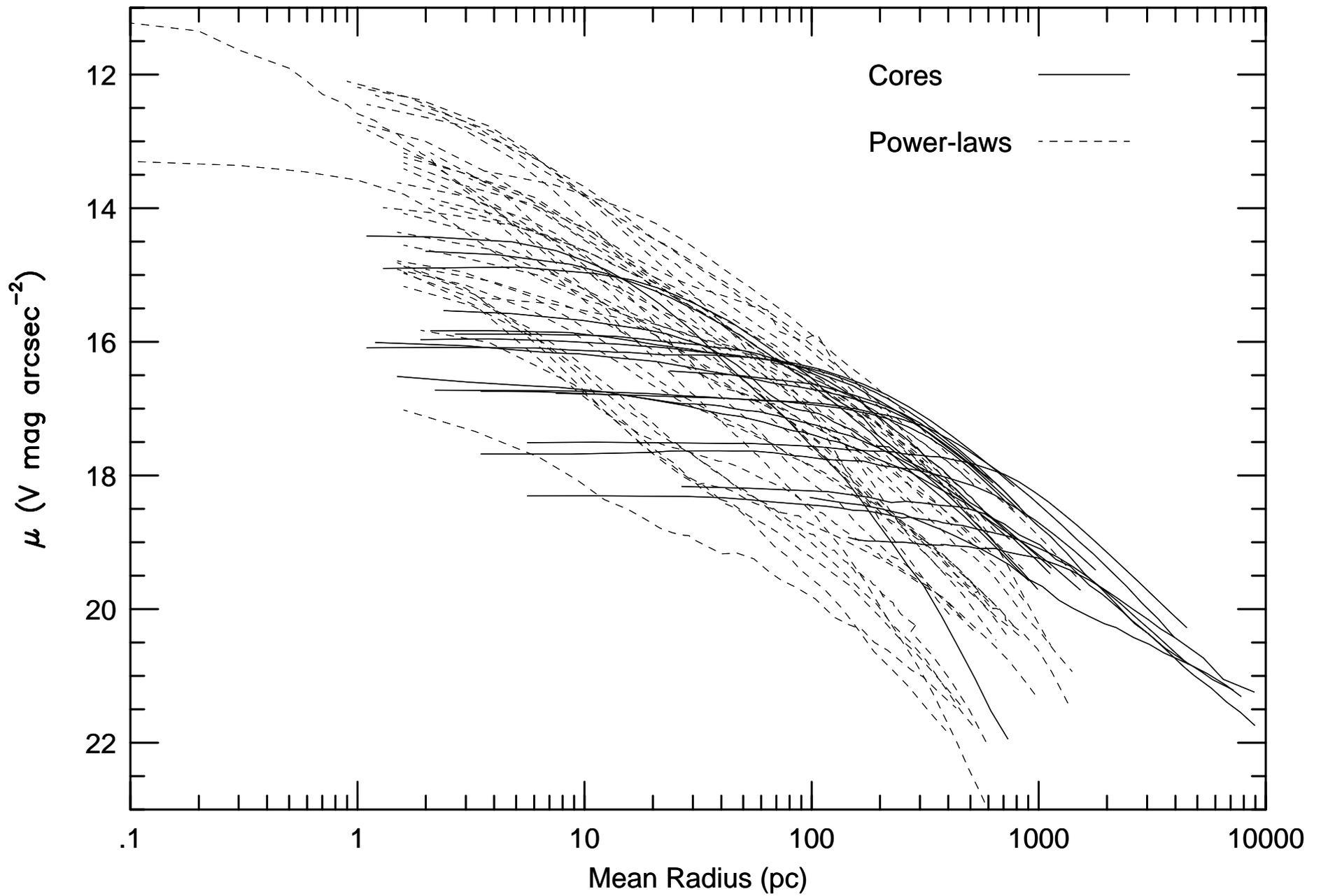
{\bf Figure 1:} V-band surface-brightness profiles of 55 ellipticals and
bulges from {\it HST}.  All were observed in the 
WFPC1 Planetary Camera through
filter F555W and were deconvolved using the Lucy-Richardson algorithm
as described in Paper I. Core galaxies (see Section 2) are plotted as solid
lines, and power-law galaxies are plotted as dashed lines. ``Mean radius'' is
the geometric mean of the semi-major and semi-minor axes of the isophotal
ellipse.

\bigskip\noindent
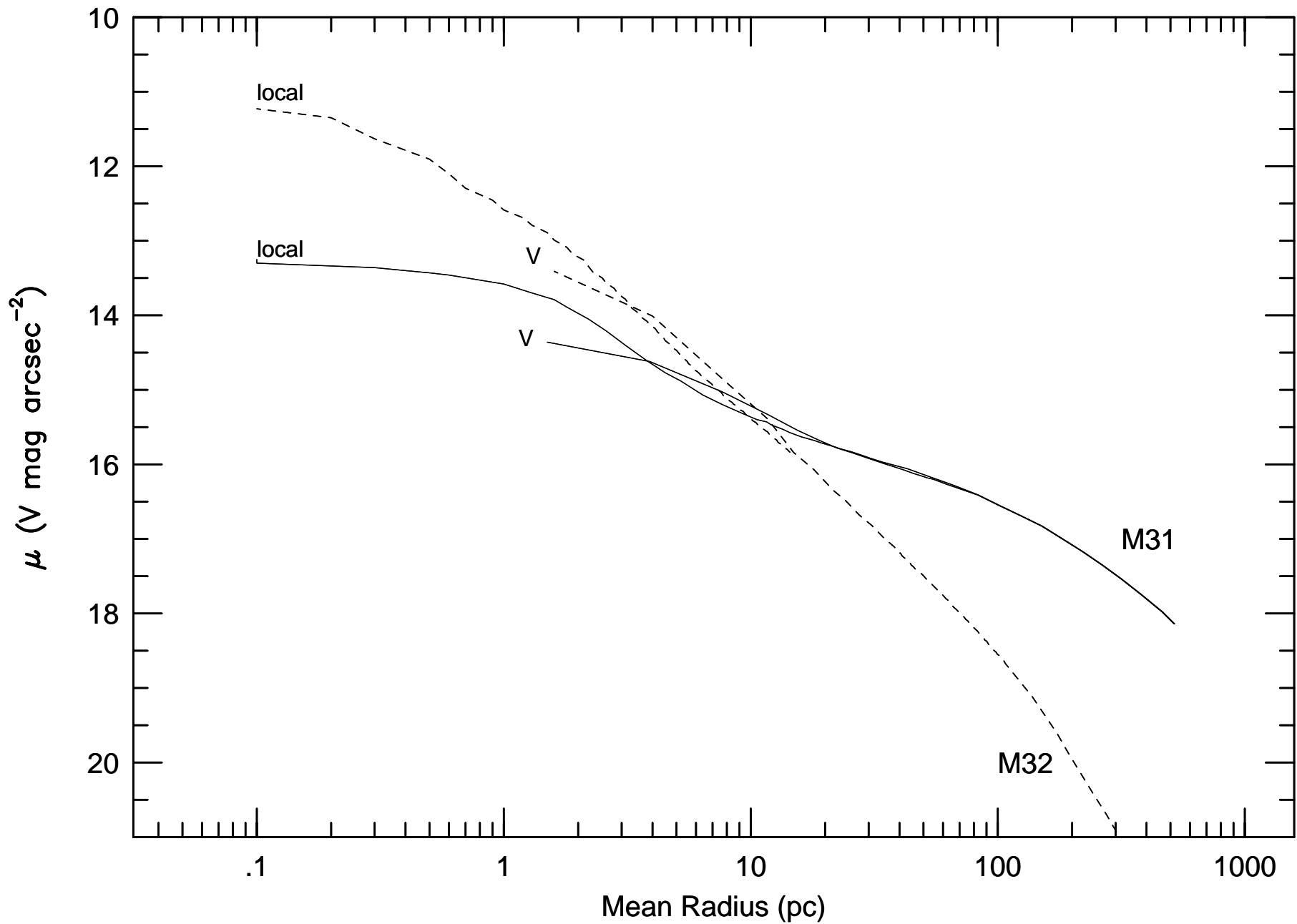
{\bf Figure 2:} {\it HST} surface-brightness profiles of M~31 and M~32, as seen
locally and near Virgo (24 times farther).  To simulate Virgo, the nearby
profile was binned by a factor of 24, convolved with the WFPC1 point-spread
function, and deconvolved with 80 iterations of the Lucy-Richardson
algorithm.  

\bigskip\noindent
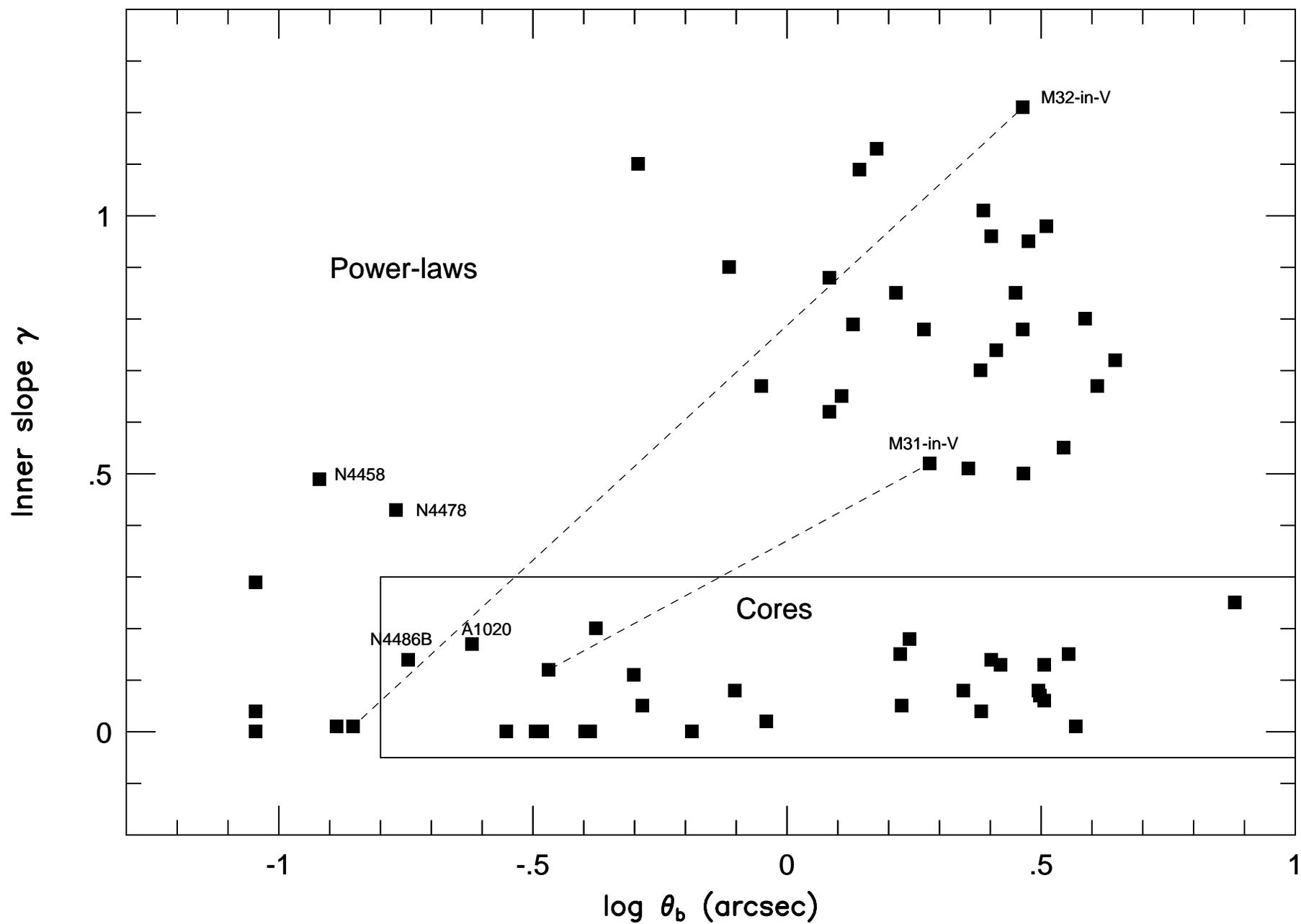
{\bf Figure 3:} Division of the sample into cores and power laws.  The figure
plots logarithmic inner slope of the surface-brightness profile, $\gamma$,
versus angular break radius, $\theta_b$, from fits to the nuker law (Eq. 2).
Galaxies with ${\rm log}~ \theta_b > -0.8$ are well resolved and divide into
two groups with high and low $\gamma$.  Dashed lines connecting the near and
far versions of M~31 and M~32 indicate potential resolution effects on other
power-law galaxies.  A galaxy must have $\gamma < 0.3$ and a well-resolved
break radius to be classed as a core.  Galaxies within the box comprise the
``Core'' sample.

\bigskip\noindent
{\bf Figure 4:} {\it HST} measurements of central parameters of hot galaxies,
as a function of absolute V magnitude.  Hubble type and nucleus types are
taken from Table 1; ``bulges'' are S0--Sb galaxies.  $r_b$ and $\mu_b$ for
power laws are limits $r^{lim}_b$ and $\mu^{lim}_b$ 
from Table 2.  M~31 and M~32
are plotted twice: asterisks show data as observed, and tails indicate their
positions as they would appear 
24 times farther away near Virgo.  The small black square is
the S0 galaxy NGC~524, which is the only core within a bulge.  The apparent
turndown in surface brightness at faint magnitudes in panel (c) is probably
a resolution effect (cf. M~32).  Effective radii are plotted in panel
(d), to be compared with break radii in panel (a): the strong impressions of
scatter at intermediate magnitudes ($-22 < M_V < -20.5$) 
and of two types of galaxies in panel
(a) are absent in panel (d).

\bigskip\noindent
{\bf Figure 5:} Break radius $r_b$ versus distance.  The dashed line is the
adopted dividing line for cores in Fig. 3 (log $\theta_b = -0.8$). 
Above this line, a core-type profile will
be seen as a resolved core, below it will be classed as a power-law.  The trend
versus distance is opposite to what one would have expected if cores and
power-laws were merely an artifact of angular resolution --- core galaxies are
on average {\it more} distant than power-laws.  Moreover, most of the sample
is close to Virgo in distance ($\log D \sim 3.2$), yet contains both cores and
power laws, confirming that the two types are intrinsically different.

\bigskip\noindent
{\bf Figure 6:} Various densities at radius 0\s.1 \  plotted against absolute
magnitude. Mass densities are derived by normalizing 
nuker-law surface-brightness fits to central $\sigma_0$.  The
symbols are the same as in Fig. 4.  Model details are given in the
text and notes to Table 3.  Panel (a) luminosity density; panel (b) mass
density; panel (c) peak Maxwellian phase-space density.  Note the range of
almost 10$^6$ in density in all three panels.  Turndowns for small galaxies
are probably an artifact of resolution (cf. M~32).

\bigskip\noindent
{\bf Figure 7a:} Replot of Fig. 4a with symbols indicating rotation speed
$(v/\sigma)_*$.  Slow rotators (filled symbols) have $(v/\sigma)_* < 0.51$;
fast rotators (open circles) have $(v/\sigma)_* \geq 0.51$.  Bulges lacking
data are classed as fast rotators. Galaxies with core profiles are indicated
by the enclosing squares; all others are power laws. The data indicate a
tendency for fast rotators to have power-law profiles.

\bigskip\noindent
{\bf Figure 7b:} Same as Fig. 7a but with symbols indicating isophotal shape
$a_4/a$.  Galaxies are classed as disky if $a_4/a \geq 0.4$, otherwise as
boxy/neutral.  Irregular profiles with variable $a_4/a$ are also classed as
boxy/neutral.  Bulges (Hubble types S0--Sb) are classed as disky. The data
indicate a tendency for disky galaxies to have power-law profiles.

\bigskip\noindent
{\bf Figure 8:} {\it HST} measurements of central parameters of hot galaxies
in fundamental-plane space.  Symbols are the same as in Fig. 4.  
Tails on M~31 and M~32 (asterisks) show the effect of moving these
galaxies 24 times further away to the vicinity of Virgo.  Resolution effects
on other power-law galaxies may be similar
and are indicated schematically by the limit flags.  Panel (d) shows the 
fundamental plane
rotated about the $\sigma_0$ axis and viewed edge-on
(for cores).  The rotation chosen
uses the same power-law combination of $r_b$ and $I_b$ used for the global
fundamental plane of elliptical galaxies by Faber \et \ (1987) and is consistent
with their best core plane within the statistical errors.  The rms scatter about
the central plane (cores only) is 0.12 dex, which is 50\% greater than the
scatter about the global plane.  This increase may be due in part to the
influence of central BHs on $\sigma_0$.  Five BH candidates from Kormendy \&
Richstone (1995) are marked in panel (d).

\bigskip\noindent
{\bf Figure 9:} Core versus global properties.  Panel (a) plots indicative core
magnitude (computed from $L_{core} \equiv \pi r_b^2 I_b $) versus total
magnitude. The dashed line is a mean fit assuming unit logarithmic slope (see
text).  Panel (b) is similar but compares break radius $r_b$ to effective
radius $r_e$.  The shaded areas represent predictions of decaying BH
binary models (Quinlan \& Hernquist 1997).

\bigskip\noindent
{\bf Figure 10:} Inner 
disk prominence versus ellipticity for power-law galaxies. Disk
score (Table 4) is a visual estimate of the evidence for an inner disk: 0 = no
evidence, 1 = slight, 2 = probable, 3 = definite.  Symbols are the same as for
power laws in Fig. 4.  Core galaxies are not plotted individually; they lie
within the rectangle.  If bulges (S0--Sb) are ignored, there is little
remaining tendency for power-law galaxies to be highly flattened, as might be
expected if they were due to inner disks seen edge-on (Jaffe \et 1994).  There
is also little tendency for inner disks to appear in flattened galaxies, which
would be expected if they were aligned with the outer isophotes.  See 
Appendix B for further discussion.

\vfill\eject\bye